\documentclass{article}
\usepackage[utf8]{inputenc}
\usepackage[margin=1.1in]{geometry}

\usepackage[numbers]{natbib}
\usepackage{xcolor}
\usepackage{url}
\usepackage{indentfirst}
\usepackage{booktabs}
\usepackage{lscape}
\usepackage{longtable}
\usepackage{ragged2e}
\usepackage{hyperref}
\usepackage{multicol}
\usepackage{enumitem}
\usepackage{amsmath}
\usepackage{dsfont}
\usepackage{authblk}

\usepackage{graphicx}

\usepackage{comment}

\newcommand{\wrt}{\textit{w.r.t. }}
\newcommand{\eg}{\textit{e.g. }}
\newcommand{\ie}{\textit{i.e. }}
\newcommand\noteg[1]{\textcolor{red}{#1}}{\tiny }
{\tiny }
\newcommand\notea[1]{\textcolor{blue}{#1}}{\tiny }

\title{Epidemic Models for COVID-19 during the First Wave from February to May 2020: a Methodological Review}

\author[1]{Marie Garin\thanks{Equal contribution. Contact: $\{$\url{ marie.garin,myrto.limnios,alice.nicolai}$\}$ \url{@ens-paris-saclay.fr}}}
\author[*1]{, Myrto Limnios}
\author[*1]{, Alice Nicola\"i}

\author[1]{, Ioannis Bargiotas}
\author[1]{, Olivier Boulant}
\author[2]{, Stephen E. Chick}
\author[1]{, Amir Dib}
\author[2]{, Theodoros Evgeniou}
\author[1]{, Mathilde Fekom}
\author[1]{, Argyris Kalogeratos}
\author[1]{, Christophe Labourdette}
\author[2,3]{, Anton Ovchinnikov}
\author[4]{, Rapha\"el Porcher}
\author[5]{, Camille Pouchol}
\author[1]{, Nicolas Vayatis}

\affil[1]{Université Paris-Saclay, ENS Paris-Saclay, CNRS, Centre Borelli, F-91190 Gif-sur-Yvette, France}
\affil[2]{INSEAD, Boulevard de Constance, 77300 Fontainebleau, France}
\affil[3]{Smith School of Business, Queen’s University, Kingston, ON, K7L3N6, Canada}
\affil[4]{Universit\'e de Paris CRESS, INSERM, INRA, 75004 Paris, France}
\affil[5]{MAP5 Laboratory, FP2M, CNRS FR 2036, Universit{\'e} de Paris, 75006 Paris, France}

\date{ }

\begin{document}

\maketitle

\begin{abstract}
    \noindent We review epidemiological models for the propagation of the COVID-19 pandemic during the early months of the outbreak: from February to May 2020. The aim is to propose a methodological review that highlights the following characteristics: $(i)$ the epidemic propagation models, $(ii)$ the modeling of intervention strategies, $(iii)$ the models and estimation procedures of the epidemic parameters and $(iv)$ the characteristics of the data used. We finally selected $80$ articles from open access databases based on criteria such as the theoretical background, the reproducibility, the incorporation of interventions strategies, \textit{etc}. It mainly resulted to phenomenological, compartmental and individual-level models. A digital companion including an online sheet, a Kibana interface and a markdown document is proposed. Finally, this work provides an opportunity to witness how the scientific community reacted to this unique situation.
\end{abstract}

\section{Introduction}

In the early months of the COVID-19 pandemic, dozens of thousands of research articles have been produced (source: \url{https://www.semanticscholar.org/cord19} with 59,888 articles referenced by May 1, 2020). The present review focuses on the subset of articles developing and/or using   mathematical models of the COVID-19 transmission during this period. In a broader context, many review articles dedicated to the modeling of diseases propagation have been published. They include, for instance, mathematical formulations of the different models to estimate epidemic parameters, forecast the epidemic or assess the impact of intervention strategies \cite{chowell2017fitting, chowell2016mathematical}. Moreover, several of these reviews offer a categorization of the different models with a precise terminology \cite{willem2017lessons, keeling2005networks, chowell2017fitting} or a mapping of the articles with their key features \cite{verelst2016behavioural, willem2017lessons}. Since November 2020, we can mention many projects of the latter type focusing on the COVID-19 epidemic, including published reviews \cite{rahimi2020review, gola2020review, gallo2020ten} and continuously updated works of scientific watch \cite{modcov19review, insermcovidreview}.\\


In the present manuscript, we review the early endeavour of mathematical modeling of the epidemic propagation. This effort of review seems crucial given the quantity and the diversity of research works produced within a short lapse of time.
In particular, the aim is to valorize this prolific production of works and to facilitate the identification of models by offering a mapping of the approaches proposed from February to 
mid-April 2020. It is completed with supplementary external contributions until the 3rd of May 2020 through our online repository. That incidentally offers the opportunity to observe the way the scientific community responded to the crisis. 
We emphasize criteria such as conceptual innovations, transparency and reproducibility of both methods and results, as well as the availability of online open-source material such as code or web demonstrations. Indeed, the ability to audit, challenge and reproduce quickly has shown to be key in order to bring the state of  knowledge closer to the settings in which policy-makers operate. Lastly, a comparison of either the numerical results or the conclusions of each article is not provided.\\




The remainder of the document is organized as follows. In Section \ref{sec:generalities}, generalities on the three main epidemic models -- phenomenological, transmission and individual-level -- and on the key epidemic parameters are provided. The Section \ref{sec:methodo} exposes the screening methodology to select the articles to be included in the review and the attributes used to describe each article are defined. In Section \ref{sec:results}, a synthetic summary of the epidemic models is proposed, as well as the methods used to estimate the parameters, the various approaches to take into account possible intervention strategies are discussed with some insights on data-related aspects. Finally, Section \ref{sec:tabular} gathers a set of tabular views with complete mapping of the reviewed articles, followed by a conclusion in Section \ref{sec:conclusion}. A digital companion of the present review including an online sheet, a Kibana interface and a markdown document is accessible on the github page: \url{https://github.com/MyrtoLimnios/covid19-biblio}.

\section{Generalities on epidemic spreading models}\label{sec:generalities}

In this section we present a synthetic view of the main characteristics and attributes of epidemic models, trying to adopt a nomenclature as faithful as possible as the one used in epidemic reviews \cite{chowell2017fitting, verelst2016behavioural, willem2017lessons, nelson2014infectious, keeling2005networks, chowell2016mathematical}. In particular, Section \ref{sec:model_cat} recalls the main model types used to characterize an epidemic, and in Section \ref{sssec:epiparam}, a list of the key epidemic parameters is defined.

\subsection{Modeling the spread of diseases}\label{sec:model_cat}

Different models for epidemics progagation exist, which differ according to the scale of analysis, complexity of the parameterization, and the practical implications of the results.

More precisely, models can be classified into two main categories: phenomenological and transmission models (see the review therein \cite{chowell2016mathematical}).

\subsubsection*{Phenomenological models}

In \textit{phenomenological models}, the curve of a time series representing the epidemic propagation (\eg time series of confirmed cases or deaths) and is assumed to have a specific shape, based on empirical data. The transmission mechanisms that give rise to the observed pattern are not explicitly modeled. For fitting and simulation, these models are usually discretized through numerical schemes. 
For instance, patterns can be estimated using (generalized) regression modeling formulation, offering a direct statistical framework to learn from empirical data.
One classical model of this type is the logistic growth model \citep{chowell2019novel} where the evolution of the number of cumulative cases at time $t$ is given by:
\begin{equation}
    \frac{dC(t)}{dt} = r C(t) \left(1-\frac{C(t)}{\text{K}}\right).
\end{equation}
where $r$ and $\text{K}$ are positive parameters. This equation characterizes a first period where the new number of cases at each step strictly increases, followed by a strict decreasing regime until the cumulative number of cases converges to its maximum value at $\text{K}$. This maximum can represent for example the total size of the population. When $C(t)$ is small with respect to $\text{K}$, the rate of growth is essentially dominated by $rC(t)$, which corresponds to an exponential growth. The growth will decrease when the population reaches the threshold of $\text{K} / 2$.

\subsubsection*{Transmission models}

\textit{Transmission} or \textit{mechanistic models} explicit the process of transmission involved in the spreading of the disease in a given population. Models of this class can be further divided in two main categories, according to their scale of analysis: \textit{compartmental models} and \textit{individual-level models}.

\medskip

\textit{Compartmental models} express the transmission dynamics at the population-level. The population is aggregated into compartments corresponding to particular health states (usually  Susceptible, Infected, Removed compartments, in the classical SIR model introduced in \cite{kermack1927contribution}). The temporal evolution of the size of each compartment is given by a system of differential equations.
In the classical SIR model for example, at time $t$ and given a population of $n$ individuals, $S(t)$ represents the number of susceptible individuals, $I(t)$ the number of infected individuals and $R(t)$ the number of removed individuals. The following equations describe the evolution of the system:
\begin{equation}\label{eq:SIR}
\left\lbrace
\begin{aligned}
\frac{dS}{dt}(t) & = -\beta S(t)I(t) \\
\frac{dI}{dt} (t) &= \beta S(t)I(t) - \gamma I(t) \\
\frac{dR}{dt}(t) &= \gamma I(t)
\end{aligned}
\right.
\end{equation}
Individuals' heterogeneity and inter-individuals' heterogeneous interactions may be modeled through the partitioning into subpopulations. Usually, numerical schemes are used to fit and/or simulate the model.


\medskip


 \textit{Individual-level models} explicitly define a state for each individual in the population. Therefore such models can incorporate more refined heterogeneity and stochasticity than population-level models, possibly at the expense of computational complexity, collecting appropriate datasets, and the increased number of parameters. As a result, parameter estimation may be difficult \cite{hazelbag2020calibration} and simulation running times may be slow compared to other models \cite{gallagher2017comparing}.
Given their high  flexibility, understanding and describing these models through a unified taxonomy is still an important open research area \cite{nepomuceno2019survey, hunter2017taxonomy}. These models are sometimes called agent-based or individual-based models.
One example of individual-level model, which is a simplified version of the model in \cite{ferguson2005strategies}, is the following. Individuals $I_1, \dots I_n$ are assigned a household in $\{h_1, \dots, h_k\}$ and can be infected inside their household (to model transmission during lockdown for instance) and a infectiousness $\rho_i$ coefficient. The probability that individual $I_i$ is infected between $t$ and $t + dt$ is given by:

\begin{equation}
    p_{i}(t, t+dt) = 1 - \exp{ \left( - \sum_{k, h_k = h_i} \int_t^{t+dt}  \lambda_k(s) ds \right) }
\end{equation}

with

\begin{equation*}
\lambda_k(s) =
\left\lbrace
\begin{aligned}
\rho_k f(s - \tau_k) & \quad \text{if individual } I_k \text{ is infected at time } s, \\
0 & \quad \text{otherwise.}
\end{aligned}
\right.
\end{equation*}
where $\tau_k$ is the time where individual $I_k$ becomes infectious and $f$ defines the infectiousness in function of the time elapsed from the end of the latent period.




\subsection{Classical epidemic parameters}\label{sssec:epiparam}

Some classical parameters are used to analyze a disease propagation by characterizing the virus' features or its spread in a given population. They are usually learned on empirical data and plugged in propagation models. We refer to the following definitions.


\begin{description}[leftmargin=0cm]

\setlength\itemsep{0.3pt}

    \item[Basic reproduction numbers $R_0$:] Nonnegative real number that quantifies the number of secondary infected cases by one individual (during his infection period), when considering all the population as susceptible. 
    The infection will spread and may become an epidemic if $R_0 > 1$ and will decline if $R_0<1$. It can broadly be quantified through the formula: $$R_0 = \beta \tau, $$ where $\beta \geq 0$ is the average of infected individuals contaminated by the infectious population by unit time and $\tau>0$ the average period of infection. Note that in the aforementioned SIR model \eqref{eq:SIR}, $\tau = 1 / \gamma$.

    \item[Effective reproduction numbers $R_e$:] Nonnegative real number that quantifies the number of secondary infected cases by one person, when considering as susceptible the current state of the population. It is time-dependent and can therefore be estimated by multiplying $R_0$ by the proportion, denoted by $s(t)$, of the susceptible population at a given time, \ie $R_e(t) = s(t) R_0 $.

    \item[Key time-to-events intervals:] Set of periods of time between each clinical state of the disease, referred to as event. In particular for the following: the \textit{incubation period} as the time-delay between exposure and the onset of clinical symptoms; the \textit{infectious period} as the time-delay of infectivity between the beginning to the end of the infection; the \textit{latent period} as the time-delay between infection and infectivity; the \textit{generation time} as the time-delay of the symptoms onset between the couple infector-infectee and the \textit{introduction date} as the date of the first infection in a fixed/given population.

    \item[Key time-to-events rates:] Set of rates relating a clinical state of the disease to another. Especially used in compartmental models to quantify the compartment's transitions (see Subsection \ref{sec:model_cat}) in a given population through the following rates: the \textit{transmission rate} from the susceptible population to the infected one; the \textit{recovery rate} from the infected population to the recovered one; \textit{mortality rate} from any given state to death.

\end{description}

\section{Review methods}\label{sec:methodo}




\subsection{Search strategy}\label{sssec:selection}


The initial step relied on a search over article bases using specific and predefined keywords. We decided to include articles regardless of their submission type and and/or status to provide an historical perspective of the proposed models. We intentionally refer to the first version available of the articles even though updated releases could exist and/or have been published.

The search strategy encompasses two main source types. First, to establish the main corpus of the review, the process was based on an extended search based on a set of keywords applied on the three main online open access archives: arXiv, biorXiv, medrXiv. Precisely, motivated by mathematical-based models and as both biorXiv and medrXiv do not provide such filters, our main focus led to the database arXiv. Nonetheless, our search on the two others came naturally through related articles citations.

The search methodology on arXiv is described as follows. We looked into the special directory "COVID-19 SARS-CoV-2 preprints" until the 11th of April 2020, with the following filters:

\begin{itemize}

    \item (``include$\_$cross$\_$list: True'') AND
    \item (``terms:''  ``title=COVID-19''  OR ``abstract=COVID-19'' OR ``abstract=SARS-CoV-2''OR \\ ``title=SARS-CoV-2'' OR
     ``title=coronavirus'' OR "abstract=coronavirus'') AND
    \item (``classification: Computer Science (cs)'' OR ``classification: Mathematics (math)'' OR ``classification: Statistics (stat)'') 
\end{itemize}



Secondly, since our reviewing effort is publicly available on github (\cite{github_review}), some other articles were identified thanks to external contributors leading to heuristic search from centers of excellence for instance. In particular, major centers collecting references and information about the epidemic were listed, such as: the MRC Centre for Global Infectious Disease Analysis Team leading the COVID-19 Response Team from Imperial College in the UK, the Institute of Health Metrics and Evaluation (IHME) from the University of Washington from the US and the Research and Action Team (REACTing) from  Inserm from France. The selected articles were added up to the the 3rd of May 2020.   



\subsection{Eligibility criteria}


Subsequently, the screening process with a more in-depth selection followed. The first batch of articles was hence reviewed based on the criteria: $(i)$ the depth of theoretical background, $(ii)$ the standards of reproducibility, $(iii)$ the originality of method and  parameters introduced, $(iv)$ the quality of exposition of the methodology, $(v)$ the capacity to encompass intervention strategies and $(vi)$ the availability of  data and code to test the model.

\subsection{Data extraction}
\label{sssec:criteria}

Once the eligible articles were selected, a data extraction protocol was used. We defined a list of categories such that it encompasses the diversity of the contributions while highlighting the similarities among them, that are listed below. Notice incidentally that these categories are not necessarily mutually exclusive and possible conflicts were solved thanks to discussion. 

\begin{itemize}
  
    \item\textit{Global approach.} General overview of the article through the following characteristics: $(i)$ estimation of epidemic parameters: based on computation or inference from data, $(ii)$ evolution forecast: prediction of future values of key indexes, \textit{e.g.} the number of infected and/or deaths, $(iii)$ modeling of various intervention strategies after governmental decisions, $(iv)$ reference to economic indicators to measure the impact of the epidemic and/or intervention strategies, $(v)$ optimization of intervention strategies (stochastic control).

    \item\textit{Data used.} Information about the nature and the source of the numerical data used.
    
    \item\textit{Model nature.} Whether the model is deterministic or stochastic.
    
    \item\textit{Model category.} Type of modeling: \textit{statistical estimation} if the model is purely statistic and not a spreading model, else one of the categories described in \ref{sec:model_cat}: either \textit{phenomenological}, \textit{compartmental} or \textit{individual-level model}. Additional attributes of the models are also reported and are specific to each category -- for instance, the difference types of compartments in compartmental models.
    

    
    
    \item\textit{Modeling of intervention strategies.} 
     How the interventions strategies are mathematically incorporated in the model:  $(i)$ addition of compartments, $(ii)$ modification of the contact matrix, $(iii)$ addition of predictive variables, $(iv)$ modification of model parameters, $(v)$ integration of strategies in the structure of the network.  
    
    
    
    

    \item\textit{Epidemic parameters.} Epidemic-related input parameters introduced in the formulation of the problem, \eg the transmission rate, the incubation period.
    
    
    
    \item\textit{Estimation method for the input parameters.} Whether parameters are inferred from a statistical framework or from the literature.

    \item\textit{Code availability.} Whether the source algorithmic code is available. 

\end{itemize}
    



In the present article we chose to describe the articles with respect to the four main attributes: the model category, the type of modeling of the intervention strategies, the estimation method for the input parameters and the data used. Nevertheless, all information is gathered at length through these categories in the online companion tools (\cite{kibana_review}, \cite{googlesheet_review}), see subsection \ref{subsec:comptool}.

\section{Results}\label{sec:results}
\medskip
The first subsection \ref{sec: search results} reports the number of articles that have been found during the search procedure. We then sequentially present a comprehensive analysis of the articles reviewed as follows. Subsection \ref{sec:models} presents the propagation models proposed in the selected papers for the disease, in particular categorized through either phenomenological, compartmental or individual-based models. In Subsection \ref{sec:interv}, the intervention strategies are summarized when introduced in the aforementioned modelings. Then the estimation methods for the set of epidemic parameters, defined in Subsection \ref{sssec:epiparam}, are discussed in Subsection \ref{sec:param}. Finally, Subsection \ref{sec:data} gathers the characteristics of the real datasets used.


\medskip
\subsection{Search results} \label{sec: search results}
\begin{figure}[h]
    \centering
    \includegraphics[height=75mm]{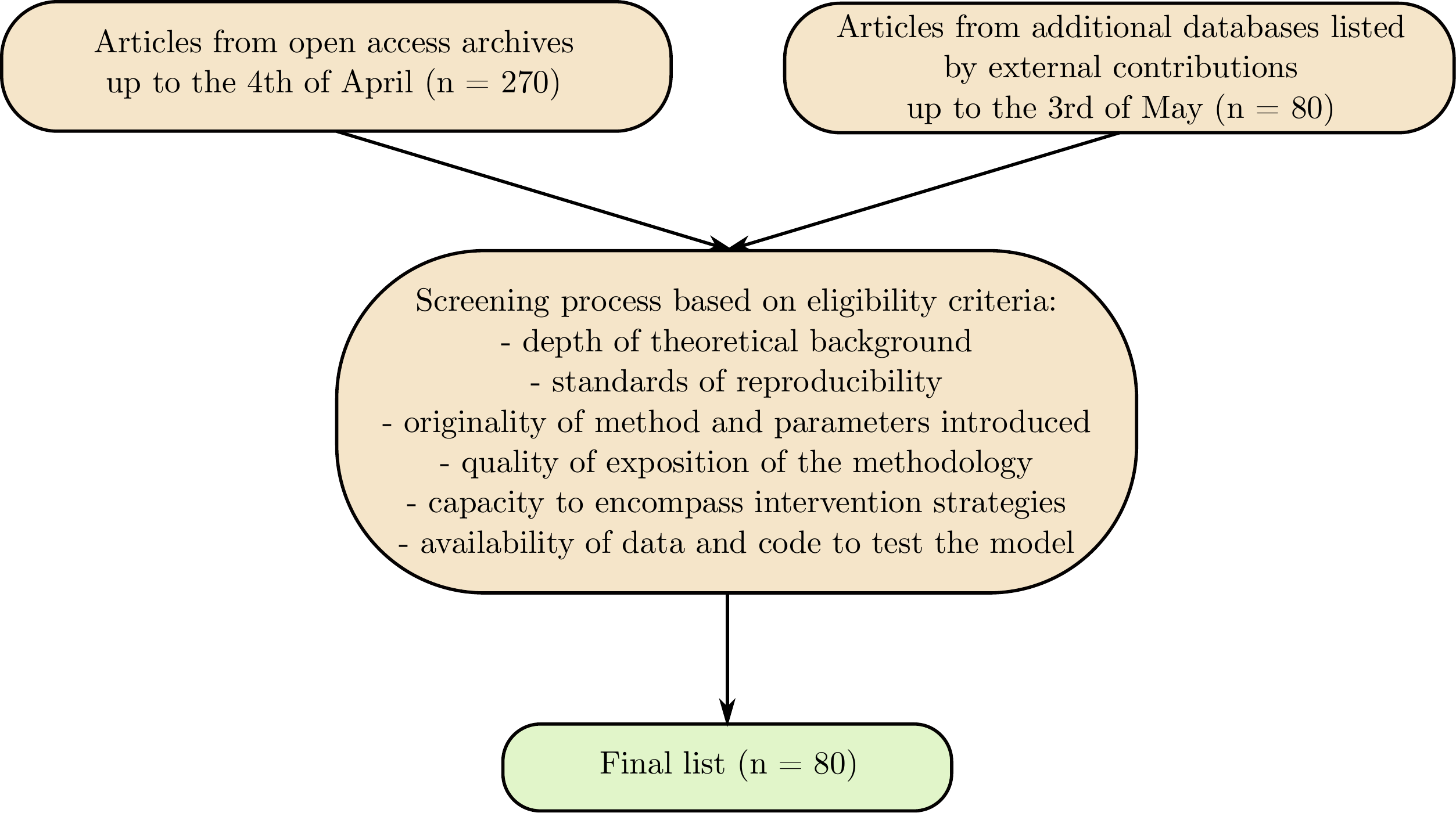}
    \caption{Diagram of the process for selecting the articles.}
    \label{fig:selection}
\end{figure}

The selection procedure resulted to a total of $41$ articles from the online open source archives, whereas $39$ were obtained thanks to external contributions.  Regarding the open source arXiv, $150$ articles were selected from the classification `Computer Science (cs)', $60$  from 'Statistics (stat)' and $25$ from `Mathematics (math)'. From both medrXiv and biorXiv, approximately $35$ were selected. Finally, we selected around a half of the external contributions. The diagram in Fig. \ref{fig:selection} illustrates the search results. \\

\subsection{Epidemic propagation models}\label{sec:models}
\medskip

A variety of different models have been used to model the spread of the virus, from phenomenological models to more explicit transmission models. We synthetically describe below the main designs found in the reviewed articles.

\medskip

\subsubsection{Phenomenological models}

\smallskip

Phenomenological approaches were used in 10 articles. Mainly, the models are based on generalized regressions of the time dependent curve of interest (\eg time series of confirmed cases or deaths) on time. Also, two of the articles derive auto-regressive models of the daily number of infections over time \cite{roy2020, kraemer2020effect}.


\medskip

Authors use S-shaped curves such as logistic or Grompertz curves \cite{villa20, verity2020estimates, yang2020rational, Batista20, covid2020forecasting, zeng2020predictions} as well as exponential curves for the early data \cite{shim2020transmission, yang2020rational} to fit a cumulative count of cases over time, and bell-shaped curves such as the ERF function \cite{zeng2020predictions} to fit the daily count over time. In some models, a Poisson or negative binomial distribution is used to model the stochasticity and the uncertainty of the prediction \cite{woody2020projections, roy2020, kraemer2020effect, shim2020transmission}. 

\medskip

In the simplest design, parameters are commonly shared by all the population and are constant over time \cite{villa20, yang2020rational}. In more complex designs, heterogeneity between states or regions is modeled through the use of mixed-effects models \cite{covid2020forecasting, woody2020projections, kraemer2020effect}. We characterize such models as \textit{spatially-structured}. \textit{Time-varying covariates} have also been introduced to account for the non-stationarity associated with the time-varying availability of tests \cite{kraemer2020effect} or changes in behavior due to the implementation of intervention strategies \cite{covid2020forecasting, woody2020projections}.


\subsubsection{Compartmental models}

\smallskip

In the vast majority of the included studies (53 articles), the classical SIR, SEIR (E: exposed) and SEIRD (D: deceased) models are adopted to analyze the spread of the virus, as well as a wide variety of extensions. In its extensions, additional health states are added. The Infected state was divided into different disease stages, \eg pre-symptomatic (before symptoms) and symptomatic (after symptoms onset) \cite{roux2020covid, diexpected, di2020expected, jia2020modeling} and refinements related to symptoms and clinical conditions are precised, \eg asymptomatic, mild or severe states \cite{djidjou2020optimal, manchein2020strong, zhangenns20, prem2020effect}. 

Moreover, hospitalization and admission to ICU compartments are introduced to predict resources needs of the healthcare system \cite{roux2020covid, di2020expected, salje2020estimating}. In order to account for the difficulty of measuring the exact size of the contaminated population, some articles divide the infected compartment into reported and unreported cases \cite{Thakkar20, calafiore2020modified, magal2020predicting}. Lastly, to model the control measures introduced to mitigate the propagation, non-working, confined or quarantine states have been introduced \cite{gollier2020policy, manchein2020strong, chen2020time}.


\paragraph{\textit{Time-delayed} and non-stationary dynamics.}
For most of the models, the dynamic of the population evolution, at a given time, only depends on the previous time step. However, some articles introduce dependence on multiple past time steps of the dynamics to account for realistic delays induced by key time-to-event variables such as the incubation period or the generation time in \cite{flaxman2020report, chen2020time, volpert2020quarantine}, see Section \ref{sssec:epiparam} for more details on these parameters. Also, non-constant parameters are introduced to capture time-varying aspects, see  \cite{kucharski2020early, chen2020scenario, zheng2020total}.

\paragraph{Stochastic models.}
A large proportion of the compartmental models, 17 articles, introduce stochasticity through stochastic transitions between compartments \cite{flaxman2020report, di2020expected, zhigljavsky2020generic, weissman2020locally, khadilkar2020optimising, simha2020simple, wu2020estimating, roques2020modele, kucharski2020early, wu2020nowcasting}, using for instance Bernoulli distributions \cite{di2020expected}, Poisson processes \cite{zhigljavsky2020generic}, or diffusion terms in the system of differential equations to account for volatility in the propagation \cite{simha2020simple}.

\paragraph{\textit{Age-structured models.} } 

In some models, the population is divided into age-stratified subpopulations with specific transition dynamics (10 articles). These distinct dynamics may be due to different disease-related characteristics, such as infectivity, susceptibility or vulnerability, see \cite{roux2020covid, salje2020estimating}. In most of the models, these age-groups are not considered equally-likely to interact. This heterogeneous mixing is modeled using contact matrices gathering the average frequencies of inter-groups contacts \cite{roux2020covid, salje2020estimating, di2020expected, chikina2020modeling, zhigljavsky2020generic, prem2020effect, zhangenns20, wu2020estimating, massonnaud2020covid, diexpected}.

\paragraph{\textit{Spatially-structured models.}}
Multi-level models have been developed to gather several cities, regions or even countries in a common evolution model (9 articles). Inter-region disparities are modeled through region-levelled epidemic parameters \cite{roux2020covid, chen2020scenario, diexpected}. Some models consider independent mixing between regions \cite{chen2020scenario, diexpected}, while others spatial interactions. These interactions may depend on the size of the populations and the distance between cities \cite{khadilkar2020optimising} or directly be measured as population flows between cities \cite{zhang2020predicting, li2020substantial, zhangenns20, zeng2020predictions}.

\paragraph{Disease-related heterogeneity.}

Other subpopulations are considered to account for individually-disease-related characteristics. Some models separate symptomatic to asymptomatic subpopulations upon infection or between infected with different degree of disease severity (10 articles) \cite{roux2020covid, salje2020estimating, di2020expected, manchein2020strong, magdon2020machine, wood2020planning, djidjou2020optimal, prem2020effect, zhangenns20, liutime}. Such models are characterized as \textit{Symptoms/severity structured}. To account for the possible significant level of non-reported infected individuals due to the absence of testing or substantial symptoms, the population can be divided into \textit{reported/unreported} groups (6 articles) \cite{Thakkar20, calafiore2020modified, magal2020predicting, li2020substantial, liu2020predicting, liu2020understanding}.

\subsubsection{Individual-level models}

\smallskip

The individual-level models analyze the transmission dynamics at the scale of each individual, considered as an entity in itself, (6 articles). This modeling can incorporate heterogeneity and stochasticity in the individual's temporal evolution. The three following types of this class are identified and ordered according to the degree of individuality in the model: \textit{branching processes}, \textit{network-based} models and \textit{individualized} models.

\paragraph{\textit{Branching processes}.} One article in the review \cite{yanev2020stochastic} introduces a two-type \textit{branching process} to model the growth of the epidemic. Infected individuals evolve separately and may independently transmit the disease according to a certain distribution. In this design, the infected population is divided into two types of infected individuals: discovered (type 1) and non-discovered (type 2). Discovered individuals do not take part in the further evolution of the process because they are isolated. Non-discovered individuals can either give rise to other non-discovered cases, be transformed into type 2 or leave the reproduction process.



\paragraph{\textit{Network-based models}.}

Two articles use random networks to simulate the propagation of the virus \cite{hurd2020covid, karin2020adaptive}. In both models, a population is assimilated to a network drawn at random, where nodes represent individuals and are connected by edges corresponding to social connections. In \cite{hurd2020covid}, individual characteristics such as the location, the gender or the age, are modeled through the nodes, where connections are sampled from a contact kernel depending on the inter-type similarity. In both articles, random heterogeneity in the host characteristics is modeled, for example through different initial immunity buffer \cite{hurd2020covid}, or different levels of infectiousness drawn from a long tailed distribution to account for \textit{super-spreader} individuals  \cite{karin2020adaptive}. At each time step, nodes are categorized according to the similar health states of the compartmental models. Then, nodes states are updated with respect to transition rates, depending on the current characteristics of the node and its connections. In \cite{hurd2020covid}, for example, each node experiences an accumulation of viral load and becomes infected when the viral load exceeds its immunity buffer. To model the COVID-19 outbreak, these studies propose distinct approaches: in \cite{karin2020adaptive} a population of a given size is simulated, whereas in \cite{hurd2020covid} a network represents a senior's residential centre inside the town, in order to simulate the vulnerability of the centre to contagion imported from the outer population.

\paragraph{\textit{Individualized models}.}

Three articles introduce models where individuals are completely unique and identifiable, in particular through their geographic localization \cite{wood2020planning, ferguson2020report, gleam2020modeling}. In this design, a population is simulated to reproduce for instance: a realistic geographical location of each individual; some realistic attributes proper to each individual, such as age or gender; some realistic contact patterns via the population distribution through households, workplaces and schools. In particular, the characterization of a global population is possible, for example, the population of the UK \cite{ferguson2020report}, the US \cite{ferguson2020report, wood2020planning} and the entire world \cite{gleam2020modeling}. Spatial interactions are represented by kernels defined between locations or by a network representing travel flows between subpopulations, centered around major transportation hubs \cite{gleam2020modeling}. These configurations require very rich data and parameterization to draw individuals, places and connections from realistic distributions. Similar to network-based models, each individual is in a particular health state at each time. At any time-step, individuals have a probability of transition between states depending on their characteristics and social interactions. Lastly, other realistic elements are modeled, such as a variable infectiousness in time for an infected individual \cite{ferguson2020report}, or the introduction of different viral strains \cite{wood2020planning}.


\medskip


\medskip




\subsection{Modeling of intervention strategies}\label{sec:interv}
\medskip
In the wake of the emergence of COVID-19, many countries responded through public measures to limit its spread. Broadly, the aim was both to contain the number of people affected by the disease and to reduce the risk of exceeding the healthcare system threshold capacity (\textit{e.g.}  to flatten the curve). To this end, various intervention strategies can be deployed: social-distancing recommendations, isolation of infected or susceptible individuals, school/university closure, global lockdown, encourage telework, mask-wearing requirements, business closures, restriction of group gatherings, random testing campaigns, air traffic suspensions, \textit{etc}. As a result, behavioural changes within the population led to a modification of the disease propagation. In this section, we will review the several approaches carried out to embed these strategies in the  models. \\

Particularly, some articles focus on assessing the impact of intervention strategies. Among others, in  \cite{anzai2020assessing}, the effects are quantified through a counterfactual model. In \cite{das2020}, authors consider the difference between predictions of the model trained before lockdown and actual data (with lockdown) to evaluate its effectiveness. The articles pertaining to this perspective do not include intervention strategies into their model. As for the agent-based ones (4 articles), social interaction patterns are inherent to the model. The degree of complexity handled by these designs has therefore enabled the modelization of interventions without the need for model modification. For this reason, we will not go into more details on the matter for both of these cases. \\

\subsubsection{Incorporation in compartmental models}

The intuitive approach considered by the vast majority of articles ($43$ of them) consists in a modification of the epidemic parameters. The parameter most likely to be impacted by social distancing is the \textit{transmission rate}. The basic reproduction number, $R_0$, depends on the transmission rate and the infectious period (often proportionally). The latter parameter is intrinsic to the disease and is hence less subject to variation. No model considers a change in the infectious period, therefore a modification of the transmission rate is equivalent to a modification of the $R_0$ (or of the $R_e$). This is why we can refer to one or the other in an equivalent way.\\

\paragraph{Scaling of the transmission rate.} The transmission rate can basically be multiplied by a constant to reflect the decrease in contacts induced by social distancing ($13$ articles). In \cite{evgeniou2020epidemic}, this constant can be either estimated or defined as a function of three parameters indicating the degree of isolation. This enables the simulation of various intervention strategies. Let recall that the emergence of the disease goes on into an outbreak according to the sign of $R_0 - 1$. Hence, the scaling of the transmission rate can therefore directly reflect the epidemic changeover. A frequently adopted procedure is to set the transmission rate to two values, before and after the lockdown, in order to grasp a behavioural response from the population to the strategies. An introduction of a factor of isolation strength per age group is embedded in \cite{salje2020estimating, zhigljavsky2020generic}. Nevertheless, in \cite{karin2020adaptive} authors propose an alternation of cycles applied to all the population. The cycles are composed of both a working period (regular transmission rate) and a lockdown/self-isolation period (modified transmission rate). The objective of this article is to find the optimal durations of these two phases. A similar approach is also proposed by \cite{simha2020simple}.\\

\paragraph{Piecewise time-dependent transmission rate.} Articles of the previous category only allow an alternation of the reproduction number through two fixed values (\textit{i.e.} regular/lockdown). The articles based on this approach ($18$ articles) tend to integrate more flexibility and complexity. In \cite{flaxman2020report}, the time-varying reproduction number is defined as the true $R_0$ (without interventions) multiplied by a function of six indicators representing non-pharmaceutical interventions. These indicators are activated when such a measure is put in place in a country. With a similar approach, in \cite{chen2020scenario}, the transmission rate is a time-varying function of non-pharmaceutical interventions time-span. Another example is provided in \cite{liutime}, where both the transmission rate and the recovering rate are functions of the time.\\
Additionally, in \cite{piguillem2020optimal}, $R_0$ is multiplied by a piecewise function, depending on both the intensity and the duration of policies. A modeling of testing policies is proposed by a time-dependent parameter of the same shape. This parameter represents the portion of the tested population from which infected individuals can be isolated. The purpose of the article is articulated on the search for an optimal combination of quarantine and testing policies.\\ 

\noindent{\textbf{Modification of the contact matrix.}} In models in which interactions between sub-populations are modeled by a contact matrix ($7$ articles), social distancing measures can directly be implemented. We refer to \cite{prem2020effect} for an estimation of age-specific and location-specific (home, work, school, other) contact matrices built under various physical distancing scenarios. Additionally, a reconstruction of the contact matrix is implemented by a combination of social distancing interventions in \cite{di2020expected}. Finally, five different contact matrices are considered in \cite{salje2020estimating}. \\

\noindent{\textbf{Addition of compartments}.} One way to account for the impact of behaviour patterns on the disease dynamics can be directly shaped by the addition of compartments ($7$ articles). In this sense, two articles divide the susceptible population into two sub-compartments: respectively working and confined in \cite{gollier2020policy} and submitted to low and high isolation recommendation populations in \cite{evgeniou2020epidemic}. Furthermore, isolation of individuals recorded as infected is one of the most widespread interventions. Some articles have therefore directly added compartments in the model. See \cite{victor2020mathematical} where the infected population is either quarantined or not. Also, \cite{Jiwei20} includes a home quarantined compartment and distinct compartments for symptomatic, asymptomatic and reported infected. \\

\noindent{\textbf{Spatially network-based models}.} For spatially structured models ($4$ articles), intervention strategies directly modify the structure itself. Indeed, in \cite{zeng2020predictions}, authors introduce a multimodel ODEs neural network. In particular, each node of the network is a compartmental model and links between layers can simulate the inter-provincial disease transmission using  mobility data. Finally, in both \cite{gleam2020modeling, chinazzi2020effect}, the world map is divided using the Voronoi method, centered on the major transportation hubs. The transmission dynamics are modeled through agent-based epidemic model for the mobility layers. In this framework, authors implemented travel restrictions by a decrease of mobility flow. \\

\subsubsection{Incorporation in phenomenological models.}

Considering the phenomenological models, only three articles derive a framework for mobility reduction. Two articles are based on regressions of the temporal curve of infections or deaths over time and contain covariates as weighted average of social-distancing metrics \cite{covid2020forecasting, woody2020projections}. In \cite{woody2020projections}, time-varying metrics capture the visiting variations in public spaces and of the time spent at home \textit{versus} at work. The last article incorporates social distancing by a time-varying scaling factor of the growth curve parameter \cite{shim2020transmission}. 

\subsubsection{Optimization of intervention strategies}

A significant part of the models aims to inform on decision making. Usually the method consists of learning retrospectively about the effects of the strategies or predicting the future propagation under different scenarios. The latter leads to a comparison of different scenarios with respect to the health cost measured by the number of deaths or the hospital saturation \cite{roux2020covid, salje2020estimating} or the economic cost induced by the lockdown \cite{gollier2020policy}. In order to automatically predict the best strategy to implement, a few articles introduced optimization frameworks. All these models are built on propagation models, compartmental \cite{alvarez2020simple, piguillem2020optimal, khadilkar2020optimising, djidjou2020optimal, toda20, wood2020planning} or individual-based \cite{wood2020planning}.

\medskip

In \cite{toda20}, the government directly controls the transmission rate and the threshold of confirmed cases by implementing its strategy, which directly minimizes the infection peak.
Three articles proposed deterministic optimal control methods to analyse how to reach the optimal trade-off between the direct economic costs and the ones implied by the healthcare system \cite{alvarez2020simple, piguillem2020optimal, djidjou2020optimal}. The optimal strategy, represented by the lockdown percentage over time  \cite{piguillem2020optimal, alvarez2020simple, djidjou2020optimal} and/or the level of testing over time \cite{piguillem2020optimal}, optimizes an objective function which integrates all the future costs.

\medskip

In the network-based framework of \cite{khadilkar2020optimising}, where nodes represent districts, the optimal strategy outputs the nodes that must be locked down each week \ie for which edges should be modified. For this purpose, nodes are assigned to a set of features (\eg the proportion of symptomatics within the district) that quantifies its current state. Also, a cost function is defined and integrates the future health and economic costs. A reinforcement learning algorithm is finally used to predict each week the best decision to take for each node, using a deep Q-network trained to predict the reward of each action given the current state of the node.

\medskip

In \cite{wood2020planning}, the strategy is optimized in the three following independent frameworks: a compartmental model, a stochastic compartmental model and an individual-based model. The result of a given strategy is binary (\eg if the proportion of infected individuals is below a certain threshold at time $t$). For the compartmental models, the strategy is defined by a controlled parameter which reduces the transmission. For the individual-based model, a strategy is defined by a a set of more refined parameters (\eg the isolation rate or the length of time a social distancing policy must be in place. The posterior probability of the controlled parameter conditionally on the success is estimated through bayesian inference. This estimation is repeated at each time step to select the best policy, conditioned on the new information.


\subsection{Models and estimation methods of the epidemic parameters}\label{sec:param}
\medskip


A vast majority of the referenced articles compute or learn the epidemic parameters implied, and if not, point  to already published studies on the topic.
 Indeed, $47$ of the articles are based on some parameters arbitrarily fixed or derived from the literature, whereas $63$ estimate a part, or the totality, of the parameters implied in the model. Also, among the $80$ articles, a large part ($53$ articles) estimate at least one epidemic parameter. Hence, this section is at first, an attempt to classify the main computation and estimation methods regarding the epidemic parameters and secondly, a focus on the key parameters introduced in Subsection \ref{sssec:epiparam}.  \\

Following Section \ref{sec:models}, the development of advanced structured models and the availability of data enable the derivation of the parameters for particular subpopulations. Precisely, specifications \wrt categorised groups are refined through: the age, the geographical location/community (household, school, \textit{etc.}) and position (region, state, \textit{etc.}), the possible hospitalization and the healthcare capacity (\eg \cite{piguillem2020optimal}), the documented/undocumented (\ie  unconfirmed and/or unreported), the ability of transmission (\eg to model \textit{super-spreader individuals}) and reciprocally the susceptibility of being infected and lastly, the scenario of intervention. It is therefore interesting to introduce covariate matrices, also known for a particular case as \textit{contact matrices}, to quantify the interrelations/interactions between each subpopulation. Readers can also refer to \cite{manchein20} where the Distance Correlation method is used to measure dependence between countries.





\subsubsection{Main data-driven estimation methods}
Most of the articles indexed in this review are based on simulations learned from clinical datasets, and when clearly mentioned, three main categories of methods are outlined through: (exact) deterministic methods ($13$ articles), estimators obtained either by descriptive statistical ($5$ articles) or by inferential methods ($42$ articles) combined with sampling techniques. Briefly, the following paragraphs give details on each of these categories.
Note that there are at least as many methodologies as papers, considering that often multiple techniques are employed.\\

First, the deterministic methods enclose the derivation of close-formed systems of equations, \eg the next-generation method \cite{victor2020mathematical,Jiwei20} or the Euler-Lotka equation \cite{yang2020rational,li2020estimation} for the computation of $R_0$. Also, six articles use deterministic optimization algorithms, mainly gradient-based ones such as gradient descent \cite{magdon2020machine,roques2020modele},  Lavenberg-Marquad \cite{chen2020time} and dubbed iterative Nelder-Mead \cite{zheng2020total} algorithms.\\ 

About the descriptive methods employed, usually when no (Bayesian) sensitivity analysis is derived, articles often resume some of the parameters by the empirical mean, without necessarily any precision on the estimation. Nonetheless, at least 5 articles detailed the descriptive techniques, \eg refer to the deterministic compartmental model \cite{salje2020estimating} for a detailed approach.\\

Along with these procedures, a majority of papers ($42$ articles) use inferential methods to estimate some parameters and build a sensitivity analysis. The main methods are inherited from regression analysis ($12$ articles), either frequentist or Bayesian, and likelihood-based formulations ($12$ articles). More precisely, regression models are derived through (log-)linear or non-linear settings. Also, likelihood-based formulations, mainly through maximum likelihood estimation, are explored ($10$ articles). Additionally, Bayesian statistical models, hierarchical or sequential, as well as Information Criteria for model selection can be found. \\

Lastly, parametric estimation of the epidemic parameters via \textit{curve fitting} are broadly explored ($6$ articles) where assumptions on the distributions are explicit, but generally the optimization criterion is not. Nevertheless, articles \cite{zhang2020evolving,piccolomiini2020monitoring} present different types of distribution patterns for some typical durations or rates. Finally, two articles propose stochastic optimization algorithms \cite{dandekar2020quantifying,de2020coronavirus}. 


\subsubsection{Key epidemic parameters}

\paragraph{Basic and effective reproduction numbers.}
The reproduction number, either $R_0$ or $R_e$, is a major epidemic parameter, which is not only used to model the spread of the virus but also the effectiveness of an intervention strategy (refer to Section \ref{sec:interv} for details). Hence, the parameter is mainly fixed from the beginning. Nevertheless, the estimation techniques are either from direct computation or from the estimation of the transmission rate. If estimated, it can be prior or posterior to the modeling of the spread of the virus. Actually, the articles from the same authors \cite{yang2020rational,li2020estimation} describe several possible methods that range from deterministic optimization to advanced Bayesian sampling algorithms.
More generally, readers can find deterministic methods computed ahead of the propagation model in  \cite{yang2020rational,chen2020time,li2020estimation,nadim2020short}; using for example the classic next-generation matrix method in \cite{Jiwei20,di2020expected,victor2020mathematical}. Random-based estimations are also derived, consider for instance regression and likelihood-based estimators in \cite{das2020prediction,chen2020time,li2020estimation}. Regarding estimation once the spread model is developed, \cite{yanev2020stochastic} proposes to compute the reproduction number by three specific statistical estimators. 
Sampling methods are derived to estimate the posterior distribution of the dynamic of $R_e$, mainly through Monte-Carlo-based algorithms, \eg \cite{ZHANGlit2020,li2020estimation}, refer also to \cite{gleam2020modeling,chinazzi2020effect} with the assumption of a uniform prior distribution on $R_0$, or a normal prior distribution in \cite{flaxman2020report}.

\paragraph{Key time-to-event intervals.}

The set of characteristic periods of time related to the virus or its propagation is mainly estimated through inferential methods. 
Undoubtedly, these techniques are of major interest in this context: simulating the distribution profile 
of the intervals (or rates in the following paragraph) from a state to another 
is key to an in-depth comprehension of the phenomenon. It also leads to sensitivity analysis. Therefore, the priors chosen for the specific epidemic time-periods are the following. The main result is that almost all the chosen priors are special cases to the gamma distribution, if not the gamma. Particularly, for the incubation time \cite{kraemer2020effect,wu2020estimating,linton2020incubation,zhang2020evolving}, the generation time \cite{ZHANGlit2020,shim2020transmission,zhang2020evolving}, and also for some of the transition times between states, depending of the article  \cite{verity2020estimates,wu2020estimating,leung2020first,zhang2020evolving}. Additionally, weibull and lognormal distributions are considered in \cite{zhang2020evolving} for the incubation, generation and onset-to-hospital times; as well as in \cite{linton2020incubation} for the first one. Also, the following time variables are supposed to follow an exponential distribution: the symptoms-to-report time \cite{leung2020first,kucharski2020early} and the  susceptible-to-infection \cite{karin2020adaptive}. In this sense, refer to \cite{verity2020estimates} for an account of various priors combined with different estimation methods for the age-structured population, authors also provide a sensitivity analysis of some parameters. 
 
\paragraph{Key time-to-event rates.}
Following the last paragraph, various approaches are used for the time-to-event key rates. Indeed, the modelings broadly range from ($i$) constant parameters, to  ($ii$) time-dependent and can also be randomised to perform sampling methods ($iii$). Note that the aforementioned assumptions are independent from a parameter to another, refer for example to \cite{zheng2020total} for an account of different models.\\ 
First, if the assumption ($i$) of constant parameters is assumed for the basic rates inherent to the models, deterministic approaches can be found for example in \cite{djidjou2020optimal,zheng2020total,nadim2020short}. Of course, empirical mean estimators are useful and often employed, \eg \cite{salje2020estimating}. Also, the work in \cite{li2020substantial} presents the impact on the epidemic spread of the undocumented infected population, quantified by advanced likelihood-based estimators. Under the assumption ($ii$), for example both compartmental models introduced in \cite{di2020expected,piccolomiini2020monitoring} fit the transmission rate dynamic to a decreasing exponential function, but also to piecewise constant and to rational functions in the latter \cite{piccolomiini2020monitoring}. Lastly, consider the framework ($iii$). 
Note that the rates are less randomised than the corresponding key periods. In \cite{karin2020adaptive}, prior distributions of both the exposure and the infection rates are supposed to follow an Erlang distribution. Nevertheless, the authors in \cite{piccolomiini2020monitoring} propose a modeling of the transmission rate through three different function types, namely piecewise constant, rational and exponential. In this sense, refer to \cite{verity2020estimates} for an account of various priors combined with different estimation methods for the age-structured population, authors also provide a sensitivity analysis of some parameters. Lastly, for more complex propagation models, refer to the individual-based method introduced in \cite{karin2020adaptive}, where a neural-network is defined such that the total infectivity of a node has a long-tailed distribution to enclose possible \textit{super-spreaders}.

\subsection{Characteristics of the data used}\label{sec:data}
\medskip

In this section, we identify the types of data used for each article. We only consider real datasets and do not report simulated data. The variety of data types in the categories used reflects the diversity of the approaches.



\paragraph{Clinical data.} Most of the articles use clinical data such as the number of reported infected or death cases and some models are validated on simulated data,. In many cases, they are used to evaluate the epidemic parameters implied in the model or to tune the key dates of the spreading dynamics (e.g. arrival, peak, end). Typical clinical datasets are the daily recorded numbers of infections, recoveries, deaths, admissions in hospital, transfers to intensive care units. Those records can be considered at different scales: region, country, world; and at different levels of details: age, gender. These data often come from either the World Health Organization, or the Johns Hopkins University or the Centers for Disease Control. More disease-related specific datasets, often gathered in hospitals, are sometimes used such as temporal recordings of viral shedding \cite{he2020temporal}, time from symptom onset and reporting, and from symptoms onsets to death. Finally, to forecast health service needs, data relating to hospital resources are used \cite{covid2020forecasting}, \eg ICU beds capacity \cite{roux2020covid,evgeniou2020epidemic}.



\paragraph{Mobility data.} In many articles, various datasets related to geographical mobility are exploited to model individual/population flows. It is of intereset as to quantify possible impact of behavioural change on the spread of the disease. A wide spectrum of sources was employed, that includes information and timing of intervention strategies \citep{chen2020scenario}, time spent at home \textit{versus} at work, changes of influence of public places \cite{woody2020projections}, day-night locations \cite{Thakkar20} and estimated reduction in mobility from GPS data \cite{covid2020forecasting}. This data often comes from SafeGraph or Baidu. Additionally, some articles deal with the effect of airline suspensions or the role played by exported cases. They involve data containing the travel history of infected people \cite{kraemer2020effect} and more generally a large volume of information related to air traffic \cite{chinazzi2020effect} (\eg from International Air Transport Association). Moreover, in \cite{Jiwei20}  meteorological data are processed to assess its impact on migration. Lastly, individual-based models required high-resolution population density data, distribution of workplace sizes \cite{ferguson2020report} or the locations of major transportation hubs \cite{chinazzi2020effect}. 


    
\paragraph{Economic data.}  Economic data were also considered in some studies in order to assess the cost of intervention strategies. Among them, there are daily costs of various interventions (tracing, testing), economic costs per (non-)fatal cases \cite{zhangenns20} or country vulnerability indexes \cite{gilbert2020preparedness,bayham2020impact}.

\subsection{Companion tools}\label{subsec:comptool}

The results are displayed in a synthetic mapping for fast access to the key articles of the literature of interest. A tabular form appeared to be a suitable support for the description of the articles 
(see \cite{googlesheet_review}). In this tabular view, each row represents the information related to one article and article attributes are then categorized through the different columns. A Kibana link (see \cite{kibana_review}) was created to enable fast search using various filters on the features of the google sheet to select the desired attributes. We aim to facilitate the identification of articles matching certain specific criteria such as the type of model, the estimation of epidemic parameters, the integration of intervention strategies, \textit{etc}. Finally the sheet tabular was written in the form of a markdown document accessible on the github page of the project (see \cite{github_review}).

\subsection{Tabular view}\label{sec:tabular}
\medskip

In this section we present a synthetic view of the global index table of the articles.  Only a subset of characteristics are displayed for the sake of clarity. The complete indexing is divided into five subtables: (1) Phenomenological models, (2) Compartmental deterministic models, (3) Compartmental stochastic models, (4) Individual-level models, (5) Statistical estimation models. The subtables (1), (2), (3), and (4) are gathered following the nomenclature adopted in Section \ref{sec:generalities}. The subtable (5) contains the articles that proposed methods of statistical estimation for epidemic parameters linked to the COVID-19, without developing a model of propagation.


\newpage 
\subsubsection{Phenomenological models}\label{subsec:tabpheno}
\RaggedRight

\centering

\setlength{\tabcolsep}{8pt}

\renewcommand{\arraystretch}{1.2}

\fontsize{7}{8}\selectfont 

\begin{longtable}{|p{0.5cm}|p{0.5cm}|p{1.2cm}|p{1.2cm}|p{2.5cm}|p{1.1cm}|p{1.6cm}|p{0.5cm}|}

\hline

Ref. & Link & Global & Stochastic & Model & Strategies & Parameters & Code \\
 &  & approach &  &  & modeling & estimation &  \\
\hline

\endhead

\cite{woody2020projections} & \href{https://www.medrxiv.org/content/10.1101/2020.04.16.20068163v2}{link} & 2 - 3 & Yes & negative binomial regression - 2 & 3 & 4 - 6 & No \\
\hline
\cite{covid2020forecasting} & \href{https://www.medrxiv.org/content/10.1101/2020.04.21.20074732v1}{link} & 1 - 2 - 3 & Yes & ERF curve - 2 & 3 & 4 - 6 & Yes \\
\hline
\cite{villa20} & \href{https://arxiv.org/pdf/2004.02406.pdf}{link} & 2 & No & Gompertz curve - logistic curve &  & 4 & No \\
\hline
\cite{roy2020} & \href{https://arxiv.org/pdf/2004.02281.pdf}{link} & 1 - 2 - 3 & Yes & Poisson auto-regressive model &  & 4 - 6 & Yes \\
\hline
\cite{verity2020estimates} & \href{https://www.thelancet.com/journals/laninf/article/PIIS1473-3099(20)30243-7/fulltext}{link} & 1 - 2 & No & logistic curve - 4 &  & 4 - 7 & Yes \\
\hline
\cite{kraemer2020effect} & \href{https://science.sciencemag.org/content/early/2020/03/25/science.abb4218}{link} & 1 - 2 - 3 & Yes & Poisson auto-regressive model - log-linear auto-regressive model - negative binomial auto-regressive model - 2 &  & 4 & No \\
\hline
\cite{shim2020transmission} & \href{https://www.ijidonline.com/article/S1201-9712(20)30150-8/fulltext}{link} & 1 - 2 - 3 & Yes & Poisson error structure - generalized growth curve & 4 & 4 & No \\
\hline
\cite{yang2020rational} & \href{https://arxiv.org/pdf/2003.05666.pdf}{link} & 2 & No & Gompertz curve - logistic curve - exponential curve &  & 4 - 6 & No \\
\hline
\cite{Batista20} & \href{https://www.researchgate.net/profile/Milan_Batista/publication/339240777_Estimation_of_the_final_size_of_coronavirus_epidemic_by_the_logistic_model/links/5e726ccba6fdcc37caf62603/Estimation-of-the-final-size-of-coronavirus-epidemic-by-the-logistic-model.pdf}{link} & 1 - 2 & No & logistic curve &  & 4 - 6 & Yes \\
\hline
\cite{zeng2020predictions} & \href{https://arxiv.org/ftp/arxiv/papers/2002/2002.04945.pdf}{link} & 2 & No & gaussian curve - Poisson curve - sigmoid curve &  & 4 - 7 & No \\
\hline 

\end{longtable}

\medskip

\RaggedRight

\begin{multicols}{2}

\raggedcolumns

\small

\textbf{Global approach}

\begin{itemize}

\setlength\itemsep{1pt}

\item [1] epidemic parameters estimation

\item [2] evolution forecast

\item [3] modeling of various intervention strategies

\item [4] introduction of economic components

\item [5] optimization of intervention strategies

\end{itemize}

\medskip

\textbf{Model}

\begin{itemize}

\setlength\itemsep{1pt}

\item [1] time-vaying covariates

\item [2] spatially-structured

\item [3] gender-structured

\item [4] age-structured

\end{itemize}

\medskip

\columnbreak

\medskip

\textbf{Intervention strategies modeling}

\begin{itemize}

\setlength\itemsep{1pt}

\item [1] addition of compartments

\item [2] modification of the contact matrix

\item [3] addition of predictor variables

\item [4] modification of model parameters

\item [5] integration of strategies in the network's structure

\end{itemize}

\medskip

\textbf{Parameters estimation methods}

\begin{itemize}

\setlength\itemsep{1pt}

\item [1] deterministic

\item [2] descriptive

\item [3] curve-fitting

\item [4] inferential

\item [5] optimization

\item [6] stochastic optimization

\item [7] deterministic optimization

\end{itemize}

\end{multicols}

\newpage
\subsubsection{Compartmental deterministic models}
\RaggedRight

\centering

\setlength{\tabcolsep}{8pt}

\renewcommand{\arraystretch}{1.2}

\fontsize{7}{8}\selectfont 

\begin{longtable}{|p{0.5cm}|p{0.5cm}|p{1.9cm}|p{3cm}|p{1cm}|p{1.6cm}|p{0.5cm}|}

\hline

Ref. & Link & Global & Model & Strategies & Parameters & Code \\
 &  & approach &  & modeling & estimation &  \\
\hline

\endhead

\cite{luo2020monitor} & \href{https://ddi.sutd.edu.sg/}{link} & 1 - 2 & SIR &  &  & Yes \\
\hline
\cite{roux2020covid} & \href{https://www.ea-reperes.com/wp-content/uploads/2020/04/ImpactConfinement-EHESP-20200322v1.pdf}{link} & 1 - 2 - 3 & SEIR H - 1 - 2 - 3 & 2 & 4 & Yes \\
\hline
\cite{salje2020estimating} & \href{https://hal-pasteur.archives-ouvertes.fr/pasteur-02548181}{link} & 2 - 3 & SEI ICU - 1 - 3 & 2 - 4 & 2 - 4 & No \\
\hline
\cite{gollier2020policy} & \href{https://www.tse-fr.eu/sites/default/files/TSE/documents/doc/by/gollier/policy-brief-deconfinement-c-gollier-avril-2020.pdf}{link} & 2 - 3 - 4 & SIRD & 1 - 4 &  & No \\
\hline
\cite{Thakkar20} & \href{https://covid.idmod.org/data/Physical_distancing_working_and_still_needed_to_prevent_COVID-19_resurgence.pdf}{link} & 1 - 2 - 3 & SEIR - 4 & 4 & 4 & No \\
\hline
\cite{manchein2020strong} & \href{https://arxiv.org/pdf/2004.00044.pdf}{link} & 2 - 3 - 5 & SEIRQ - 3 & 1 - 4 & 1 & No \\
\hline
\cite{leung2020first} & \href{https://www.sciencedirect.com/science/article/pii/S0140673620307467?casa_token=2RgBVkfxteIAAAAA:gbHVp1xLZAp9hYxQ2-MRtg6MIWA4WQffXkXRONqtan7NnDcdaQa14uZ_No-L1Mem500PnFuQYg}{link} & 1 - 2 - 3 & SIR & 4 & 4 & Yes \\
\hline
\cite{chikina2020modeling} & \href{https://arxiv.org/abs/2004.04144}{link} & 2 - 3 & SIRD - 1 & 2 &  & Yes \\
\hline
\cite{das2020prediction} & \href{https://arxiv.org/pdf/2004.03147.pdf}{link} & 1 - 2 & SIR &  & 4 - 7 & Yes \\
\hline
\cite{chen2020scenario} & \href{https://arxiv.org/pdf/2004.04529.pdf}{link} & 2 - 3 & SIR - 2 & 4 & 4 & No \\
\hline
\cite{alvarez2020simple} & \href{https://bfi.uchicago.edu/wp-content/uploads/BFI_WP_202034.pdf}{link} & 2 - 3 - 4 - 5 & SIR & 4 & 2 & No \\
\hline
\cite{magdon2020machine} & \href{https://arxiv.org/pdf/2003.07602.pdf}{link} & 1 - 2 & SIR - 3 & 4 & 4 - 7 & No \\
\hline
\cite{de2020coronavirus} & \href{https://arxiv.org/pdf/2004.00553.pdf}{link} & 1 - 2 - 3 & SEIRD & 4 & 4 - 6 & No \\
\hline
\cite{wood2020planning} & \href{https://arxiv.org/pdf/2003.13221.pdf}{link} & 2 - 3 - 5 & SEIR - 3 & 4 &  & Yes \\
\hline
\cite{karin2020adaptive} & \href{https://www.medrxiv.org/content/10.1101/2020.04.04.20053579v1.full.pdf}{link} & 2 - 3 & SEIR - SEIR-Erlang & 4 &  & Yes \\
\hline
\cite{piccolomiini2020monitoring} & \href{https://www.medrxiv.org/content/10.1101/2020.04.03.20049734v1.full.pdf}{link} & 1 - 2 - 3 & SEIRD & 4 & 3 & Yes \\
\hline
\cite{djidjou2020optimal} & \href{https://www.medrxiv.org/content/medrxiv/early/2020/05/15/2020.04.02.20049189.full.pdf}{link} & 1 - 2 - 3 - 4 - 5 & SEAIR - 3 & 4 & 1 - 3 & No \\
\hline
\cite{zhang2020predicting} & \href{https://papers.ssrn.com/sol3/papers.cfm?abstract_id=3559554}{link} & 2 - 3 & SICRD - 2 & 4 &  & Yes \\
\hline
\cite{calafiore2020modified} & \href{https://arxiv.org/pdf/2003.14391.pdf}{link} & 1 - 2 & SIRD - 4 &  & 4 - 7 & Yes \\
\hline
\cite{toda20} & \href{https://arxiv.org/pdf/2003.11221.pdf}{link} & 1 - 2 - 4 - 5 & SIR  & 4 &  & Yes \\
\hline
\cite{piguillem2020optimal} & \href{http://www.eief.it/eief/images/WP_20.04.pdf}{link} & 1 - 2 - 4 - 5 & SEIR & 4 & 2 & No \\
\hline
\cite{prem2020effect} & \href{https://www.sciencedirect.com/science/article/pii/S2468266720300736#sec1}{link} & 2 - 3 & SEIR - 1 - 3 & 2 &  & Yes \\
\hline
\cite{zhangenns20} & \href{https://papers.ssrn.com/sol3/papers.cfm?abstract_id=3558339}{link} & 1 - 2 - 3 - 4 & SEIR - 1 - 2 - 3 & 2 & 1 - 4 & No \\
\hline
\cite{volpert2020quarantine} & \href{https://arxiv.org/pdf/2003.09444.pdf}{link} & 2 - 3 & SIR &  & 1 & No \\
\hline
\cite{magal2020predicting} & \href{https://www.medrxiv.org/content/10.1101/2020.03.21.20040154v1}{link} & 1 - 2 - 3 & SIRU - 4 & 4 &  & No \\
\hline
\cite{victor2020mathematical} & \href{https://www.researchgate.net/publication/339944210_MATHEMATICAL_PREDICTIONS_FOR_COVID-19_AS_A_GLOBAL_PANDEMIC}{link} & 1 - 3 & SEIRUS & 1 & 1 & No \\
\hline
\cite{nadim2020short} & \href{https://arxiv.org/pdf/2003.08150.pdf}{link} & 1 - 2 - 3 & SEIRQAJ & 1 & 1 - 4 & No \\
\hline
\cite{massonnaud2020covid} & \href{https://www.medrxiv.org/content/medrxiv/early/2020/03/20/2020.03.16.20036939.full.pdf}{link} & 1 - 2 - 3 & SEIR - 1 & 4 & 2 - 4 & Yes \\
\hline
\cite{yang2020rational} & \href{https://arxiv.org/pdf/2003.05666.pdf}{link} & 2 & SIR - SEIR - SEIR-QD - SEIR-AHQ - SEIR-PO &  & 4 & No \\
\hline
\cite{Jiwei20} & \href{https://arxiv.org/pdf/2003.02985.pdf}{link} & 1 - 2 - 3 - 4 & SEIRDAQ & 1 & 1 - 2 - 4 & No \\
\hline
\cite{liutime} & \href{http://gibbs1.ee.nthu.edu.tw/A_TIME_DEPENDENT_SIR_MODEL_FOR_COVID_19.PDF}{link} & 1 - 2 - 3 & SIR - 3 & 4 & 4 & Yes \\
\hline
\cite{liu2020predicting} & \href{https://arxiv.org/pdf/2002.12298.pdf}{link} & 1 - 2 - 3 & SIRU - 4 & 4 &  & No \\
\hline
\cite{Batista20_2} & \href{https://www.researchgate.net/publication/339311383_Estimation_of_the_final_size_of_the_coronavirus_epidemic_by_the_SIR_model}{link} & 1 - 2 & SIR &  &  & Yes \\
\hline
\cite{chen2020time} & \href{https://arxiv.org/pdf/2002.02590.pdf}{link} & 1 - 2 - 3 & IRGJ & 1 & 4 - 7 & No \\
\hline
\cite{liu2020understanding} & \href{https://www.preprints.org/manuscript/202002.0079/v1}{link} & 1 - 2 - 3 & SIRU - 4 & 4 &  & No \\
\hline 

\end{longtable}

\medskip

\RaggedRight

\begin{multicols}{2}

\raggedcolumns

\small

\textbf{Global approach}

\begin{itemize}

\setlength\itemsep{1pt}

\item [1] epidemic parameters estimation

\item [2] evolution forecast

\item [3] modeling of various intervention strategies

\item [4] introduction of economic components

\item [5] optimization of intervention strategies

\end{itemize}

\medskip

\textbf{Model}

\begin{itemize}

\setlength\itemsep{1pt}

\item [1] age-structured

\item [2] spatially-structured

\item [3] symptoms/severity structured

\item [4] reported/unreported structured

\end{itemize}

\medskip

\columnbreak

\medskip

\textbf{Intervention strategies modeling}

\begin{itemize}

\setlength\itemsep{1pt}

\item [1] addition of compartments

\item [2] modification of the contact matrix

\item [3] addition of predictor variables

\item [4] modification of model parameters

\item [5] integration of strategies in the network's structure

\end{itemize}

\medskip

\textbf{Epidemic parameters estimation methods}

\begin{itemize}

\setlength\itemsep{1pt}

\item [1] deterministic

\item [2] descriptive

\item [3] curve-fitting

\item [4] inferential

\item [5] optimization

\item [6] stochastic optimization

\item [7] deterministic optimization

\end{itemize}

\end{multicols}

\newpage
\subsubsection{Compartmental stochastic models}
\RaggedRight

\centering

\setlength{\tabcolsep}{8pt}

\renewcommand{\arraystretch}{1.2}

\fontsize{7}{8}\selectfont 

\begin{longtable}{|p{0.5cm}|p{0.5cm}|p{1.9cm}|p{3cm}|p{1cm}|p{1.6cm}|p{0.5cm}|}

\hline

Ref. & Link & Global & Model & Strategies & Parameters & Code \\
 &  & approach &  & modeling & estimation &  \\
\hline

\endhead

\cite{di2020expected} & \href{https://www.epicx-lab.com/uploads/9/6/9/4/9694133/inserm-covid-19_report_lockdown_idf-20200412.pdf}{link} & 1 - 2 - 3 & SEIRD H ICU - 1 - 3 & 2 & 1 & No \\
\hline
\cite{zhigljavsky2020generic} & \href{https://arxiv.org/pdf/2004.01991.pdf}{link} & 1 - 2 - 3 & SIR - 1 & 4 & 4 & Yes \\
\hline
\cite{weissman2020locally} & \href{https://annals.org/aim/fullarticle/2764423/locally-informed-simulation-predict-hospital-capacity-needs-during-covid-19}{link} & 2 - 3 & SIR & 4 &  & Yes \\
\hline
\cite{dandekar2020quantifying} & \href{https://www.medrxiv.org/content/10.1101/2020.04.03.20052084v1.full.pdf}{link} & 1 - 2 - 3 & SIR - SEIR - SEIRQ & 1 - 4 & 4 - 6 & No \\
\hline
\cite{zheng2020total} & \href{https://arxiv.org/pdf/2004.00412.pdf}{link} & 1 - 2 - 3 & SIR - SIRQ & 1 - 4 & 1 - 4 & Yes \\
\hline
\cite{khadilkar2020optimising} & \href{https://arxiv.org/pdf/2003.14093.pdf}{link} & 3 - 5 & SEIRD - 2 & 5 &  & No \\
\hline
\cite{flaxman2020report} & \href{https://spiral.imperial.ac.uk/handle/10044/1/77731}{link} & 1 - 2 - 3 & ID & 4 & 4 & Yes \\
\hline
\cite{simha2020simple} & \href{https://arxiv.org/pdf/2003.11920.pdf}{link} & 1 - 2 - 3 & SIR & 4 &  & No \\
\hline
\cite{roques2020modele} & \href{https://arxiv.org/pdf/2003.10720.pdf}{link} & 2 & SIR &  & 4 - 7 & No \\
\hline
\cite{wu2020estimating} & \href{https://www.nature.com/articles/s41591-020-0822-7}{link} & 1 - 3 & SIR - 1 & 4 & 4 & Yes \\
\hline
\cite{li2020substantial} & \href{https://science.sciencemag.org/content/early/2020/03/24/science.abb3221}{link} & 1 - 2 & SEI - 2 - 4 &  & 4 & Yes \\
\hline
\cite{diexpected} & \href{https://www.epicx-lab.com/uploads/9/6/9/4/9694133/inserm_covid-19-school-closure-french-regions_20200313.pdf}{link} & 1 - 2 - 3 & SEIR - 1 - 2 & 2 & 2 - 3 - 4 & No \\
\hline
\cite{kucharski2020early} & \href{https://www.thelancet.com/journals/laninf/article/PIIS1473-3099(20)30144-4/fulltext}{link} & 1 - 2 & SIER & 4 & 4 & Yes \\
\hline
\cite{zeng2020predictions} & \href{https://arxiv.org/ftp/arxiv/papers/2002/2002.04945.pdf}{link} & 1 - 2 - 3 & SEIRSD & 5 & 1 & No \\
\hline
\cite{wu2020nowcasting} & \href{https://www.thelancet.com/action/showPdf?pii=S0140-6736%2820%2930260-9}{link} & 1 - 2 - 3 & SEIR & 4 & 4 & No \\
\hline 

\end{longtable}

\medskip

\RaggedRight

\begin{multicols}{2}

\raggedcolumns

\small

\textbf{Global approach}

\begin{itemize}

\setlength\itemsep{1pt}

\item [1] epidemic parameters estimation

\item [2] evolution forecast

\item [3] modeling of various intervention strategies

\item [4] introduction of economic components

\item [5] optimization of intervention strategies

\end{itemize}

\medskip

\textbf{Model}

\begin{itemize}

\setlength\itemsep{1pt}

\item [1] age-structured

\item [2] spatially-structured

\item [3] symptoms/severity structured

\item [4] reported/unreported structured

\end{itemize}

\medskip

\columnbreak

\medskip

\textbf{Intervention strategies modeling}

\begin{itemize}

\setlength\itemsep{1pt}

\item [1] addition of compartments

\item [2] modification of the contact matrix

\item [3] addition of predictor variables

\item [4] modification of model parameters

\item [5] integration of strategies in the network's structure

\end{itemize}

\medskip

\textbf{Epidemic parameters estimation methods}

\begin{itemize}

\setlength\itemsep{1pt}

\item [1] deterministic

\item [2] descriptive

\item [3] curve-fitting

\item [4] inferential

\item [5] optimization

\item [6] stochastic optimization

\item [7] deterministic optimization

\end{itemize}

\end{multicols}

\newpage
\subsubsection{Individual-level models}
\RaggedRight

\centering

\setlength{\tabcolsep}{8pt}

\renewcommand{\arraystretch}{1.2}

\fontsize{7}{8}\selectfont 

\begin{longtable}{|p{0.5cm}|p{0.5cm}|p{1.2cm}|p{1.4cm}|p{1.8cm}|p{1.2cm}|p{1.6cm}|p{0.5cm}|}

\hline

Ref. & Link & Global & Stochastic & Model & Strategies & Parameters & Code \\
 &  & approach &  &  & modeling & estimation &  \\
\hline

\endhead

\cite{hurd2020covid} & \href{https://arxiv.org/pdf/2004.02779.pdf}{link} & 2 - 3 & Yes & random network & 5 &  & No \\
\hline
\cite{wood2020planning} & \href{https://arxiv.org/pdf/2003.13221.pdf}{link} & 2 - 3 - 5 & Yes & individualized & 5 &  & Yes \\
\hline
\cite{karin2020adaptive} & \href{https://www.medrxiv.org/content/10.1101/2020.04.04.20053579v1.full.pdf}{link} & 2 - 3 & Yes & random network & 4 - 5 & 4 & Yes \\
\hline
\cite{yanev2020stochastic} & \href{https://arxiv.org/pdf/2004.00941.pdf}{link} & 2 - 3 & Yes & branching process &  &  & Yes \\
\hline
\cite{ferguson2020report} & \href{https://www.imperial.ac.uk/media/imperial-college/medicine/sph/ide/gida-fellowships/Imperial-College-COVID19-NPI-modelling-16-03-2020.pdf}{link} & 2 - 3 & Yes & individualized & 5 &  & No \\
\hline
\cite{gleam2020modeling} & \href{https://uploads-ssl.webflow.com/58e6558acc00ee8e4536c1f5/5e8bab44f5baae4c1c2a75d2_GLEAM_web.pdf}{link} & 1 - 2 - 3 & Yes & individualized & 5 & 4 & No \\
\hline
\cite{chinazzi2020effect} & \href{https://science.sciencemag.org/content/early/2020/03/05/science.aba9757}{link} & 1 - 2 - 3 & Yes & individualized & 5 & 4 & No \\
\hline 

\end{longtable}

\medskip

\RaggedRight

\begin{multicols}{2}

\raggedcolumns

\small

\textbf{Global approach}

\begin{itemize}

\setlength\itemsep{1pt}

\item [1] epidemic parameters estimation

\item [2] evolution forecast

\item [3] modeling of various intervention strategies

\item [4] introduction of economic components

\item [5] optimization of intervention strategies

\end{itemize}

\medskip

\columnbreak

\medskip

\textbf{Intervention strategies modeling}

\begin{itemize}

\setlength\itemsep{1pt}

\item [1] addition of compartments

\item [2] modification of the contact matrix

\item [3] addition of predictor variables

\item [4] modification of model parameters

\item [5] integration of strategies in the network's structure

\end{itemize}

\medskip

\textbf{Epidemic parameters estimation methods}

\begin{itemize}

\setlength\itemsep{1pt}

\item [1] deterministic

\item [2] descriptive

\item [3] curve-fitting

\item [4] inferential

\item [5] optimization

\item [6] stochastic optimization

\item [7] deterministic optimization

\end{itemize}

\end{multicols}


\newpage
\subsubsection{Statistical estimation models}
\RaggedRight

\centering

\setlength{\tabcolsep}{8pt}

\renewcommand{\arraystretch}{1.2}

\fontsize{7}{8}\selectfont 

\begin{longtable}{|p{0.5cm}|p{0.5cm}|p{1.2cm}|p{1.4cm}|p{1.6cm}|p{0.5cm}|}

\hline

Ref. & Link & Global & Stochastic & Parameters & Code \\
 &  & approach &  & estimation &  \\
\hline

\endhead

\cite{he2020temporal} & \href{https://www.nature.com/articles/s41591-020-0869-5}{link} & 1 & Yes &  & Yes \\
\hline
\cite{yanev2020stochastic} & \href{https://arxiv.org/pdf/2004.00941.pdf}{link} & 1 & Yes & 4 & Yes \\
\hline
\cite{zhang2020evolving} & \href{https://www.thelancet.com/journals/laninf/article/PIIS1473-3099(20)30230-9/fulltext}{link} & 1 - 2 - 3 & Yes & 3 - 4 & Yes \\
\hline
\cite{fong2020} & \href{https://arxiv.org/ftp/arxiv/papers/2003/2003.09868.pdf}{link} & 1 - 2 - 4 & Yes & 4  & No \\
\hline
\cite{yang2020rational} & \href{https://arxiv.org/pdf/2003.05666.pdf}{link} & 1 & Yes & 1 - 4 & Yes \\
\hline
\cite{adiga2020spread} & \href{https://www.medrxiv.org/content/medrxiv/early/2020/03/02/2020.02.20.20025882.full.pdf}{link} & 2 & No & 4 & Yes \\
\hline
\cite{adiga2020evaluating} & \href{https://www.medrxiv.org/content/medrxiv/early/2020/02/23/2020.02.20.20025882.full.pdf}{link} & 2 & No & 4 & Yes \\
\hline
\cite{li2020estimation} & \href{https://papers.ssrn.com/sol3/papers.cfm?abstract_id=3542150}{link} & 1 - 2 - 3 & Yes & 1 - 2 - 4 & Yes \\
\hline
\cite{gilbert2020preparedness} & \href{https://www.thelancet.com/journals/lancet/article/PIIS0140-6736(20)30411-6/fulltext#sec1}{link} & 1 & No & 4 & No \\
\hline
\cite{linton2020incubation} & \href{https://www.mdpi.com/2077-0383/9/2/538}{link} & 1 & Yes & 4 & Yes \\
\hline
\cite{anzai2020assessing} & \href{https://www.mdpi.com/2077-0383/9/2/601}{link} & 2 & Yes & 4 & No \\
\hline 

\end{longtable}

\medskip

\RaggedRight

\begin{multicols}{2}

\raggedcolumns

\small

\textbf{Global approach}

\begin{itemize}

\setlength\itemsep{1pt}

\item [1] epidemic parameters estimation

\item [2] evolution forecast

\item [3] modeling of various intervention strategies

\item [4] introduction of economic components

\item [5] optimization of intervention strategies

\end{itemize}

\medskip

\columnbreak

\medskip

\textbf{Intervention strategies modeling}

\begin{itemize}

\setlength\itemsep{1pt}

\item [1] addition of compartments

\item [2] modification of the contact matrix

\item [3] addition of predictor variables

\item [4] modification of model parameters

\item [5] integration of strategies in the network's structure

\end{itemize}

\medskip

\textbf{Epidemic parameters estimation methods}

\begin{itemize}

\setlength\itemsep{1pt}

\item [1] deterministic

\item [2] descriptive

\item [3] curve-fitting

\item [4] inferential

\item [5] optimization

\item [6] stochastic optimization

\item [7] deterministic optimization

\end{itemize}

\end{multicols}

\normalsize

\justify

\newpage
\section{Conclusion} \label{sec:conclusion}
This manuscript reports the modeling choices of several international teams to respond to the urgency of the first months of the outbreak. It transcripts the amazing ability of the scientific community from different fields to react to such a globalized and unprecedented event with so many diverse approaches. Indeed, we highlight the tremendous amount of new publications in such a short time-period (2-3 months) as well as the capacity of the scientists to integrate problematics from various origins (\textit{e.g.} political, societal, economical, \textit{etc.}). Also, some papers that were published online kept the results updated with respect to the latest data or knowledge of the virus. 
In particular, both modelings and forecasts were so much up-to-date that specificities of the virus were immediately taken into account, such as the important proportion of asymptomatic cases or the heterogeneity between regions.  The same holds for the particularities of the outbreak according to the resources and political decisions, e.g. the possible difficulty to test the population or monitor the diffusion networks, and the resulting strong uncertainty regarding the number of contaminations. Finally, all the debates that were being held in the society, crucial to the management of the epidemic, were in-detail examined with many realistic elements, for instance the efficiency of wearing masks and the effects of various intervention strategies. In many countries, we saw the importance of these publications for the policies that were adopted as many governments were advised, at a certain level, by scientific communities.\\

Of course, we need to shade this explosion of articles by a weakened quality regarding the following aspects. Briefly, some standard article components were sometimes poorly developed or even lacked, \textit{e.g.} the related-work section, the explanation of the modeling choices and their interpretation, the simulations and associated statistical analysis such as the sensitivity analysis, a discussion section contrasting the results, \textit{etc.}. We note that some authors acknowledged these possible shortcomings through a disclaimer at the beginning of their article. It actually questions the compatibility between the urgency of political action and the necessary hindsight inherent to the scientific research.\\
%

Finally, we would like to address the global issue related to the accessibility, the aggregation and the comparison of the data in different countries. Indeed, data acquisition was and is still a region/country-dependent process that evolves in time. Even though the geographical provenance of the data or the \textit{a-priori} knowledge on the parameters were clearly mentioned, the implication of deriving results for a different region were rarely discussed. Especially as it could entail possible misinterpretations or inappropriate uses of the numerical forecasts. 
To conclude on a positive note, we would like to highlight the prolific production of open source articles, blogs and codes that were provided and helped collective and reproducible projects in such a critical time.

\section*{Acknowledgment}

This work was supported by a public grant as part of the Investissement d'avenir project, reference ANR-11-LABX-0056-LMH, LabEx LMH, and the Région Ile-de-France.


\bibliographystyle{unsrtnat}
\bibliography{references}

\begin{thebibliography}{99}
\providecommand{\natexlab}[1]{#1}
\providecommand{\url}[1]{\texttt{#1}}
\expandafter\ifx\csname urlstyle\endcsname\relax
  \providecommand{\doi}[1]{doi: #1}\else
  \providecommand{\doi}{doi: \begingroup \urlstyle{rm}\Url}\fi

\bibitem[Chowell(2017)]{chowell2017fitting}
Gerardo Chowell.
\newblock Fitting dynamic models to epidemic outbreaks with quantified
  uncertainty: a primer for parameter uncertainty, identifiability, and
  forecasts.
\newblock \emph{Infectious Disease Modelling}, 2\penalty0 (3):\penalty0
  379--398, 2017.

\bibitem[Chowell et~al.(2016)Chowell, Sattenspiel, Bansal, and
  Viboud]{chowell2016mathematical}
Gerardo Chowell, Lisa Sattenspiel, Shweta Bansal, and C{\'e}cile Viboud.
\newblock Mathematical models to characterize early epidemic growth: A review.
\newblock \emph{Physics of life reviews}, 18:\penalty0 66--97, 2016.

\bibitem[Willem et~al.(2017)Willem, Verelst, Bilcke, Hens, and
  Beutels]{willem2017lessons}
Lander Willem, Frederik Verelst, Joke Bilcke, Niel Hens, and Philippe Beutels.
\newblock Lessons from a decade of individual-based models for infectious
  disease transmission: a systematic review (2006-2015).
\newblock \emph{BMC infectious diseases}, 17\penalty0 (1):\penalty0 612, 2017.

\bibitem[Keeling and Eames(2005)]{keeling2005networks}
Matt~J Keeling and Ken~TD Eames.
\newblock Networks and epidemic models.
\newblock \emph{Journal of the Royal Society Interface}, 2\penalty0
  (4):\penalty0 295--307, 2005.

\bibitem[Verelst et~al.(2016)Verelst, Willem, and
  Beutels]{verelst2016behavioural}
Frederik Verelst, Lander Willem, and Philippe Beutels.
\newblock Behavioural change models for infectious disease transmission: a
  systematic review (2010--2015).
\newblock \emph{Journal of The Royal Society Interface}, 13\penalty0
  (125):\penalty0 20160820, 2016.

\bibitem[Rahimi et~al.(2021)Rahimi, Chen, and Gandomi]{rahimi2020review}
Iman Rahimi, Fang Chen, and Amir~H Gandomi.
\newblock A review on covid-19 forecasting models.
\newblock \emph{Neural Computing and Applications}, pages 1--11, 2021.

\bibitem[Gola et~al.(2020)Gola, Arya, Dugh, et~al.]{gola2020review}
Abhinav Gola, Ravi~Kumar Arya, Ravi Dugh, et~al.
\newblock Review of forecasting models for coronavirus (covid-19) pandemic in
  india during country-wise lockdown.
\newblock \emph{medRxiv 2020.08.03.20167254}, 2020.

\bibitem[Gallo et~al.(2020)Gallo, de~Morais~Oliveira, Abrah{\~a}o, Sandoval,
  Martins, Almir{\'o}n, Dos~Santos, Ara{\'u}jo, de~Oliveira, and
  Peixoto]{gallo2020ten}
Luciana~Guerra Gallo, Ana~Fl{\'a}via de~Morais~Oliveira, Amanda~Amaral
  Abrah{\~a}o, Leticia Assad~Maia Sandoval, Yure Rodrigues~Ara{\'u}jo Martins,
  Maria Almir{\'o}n, Fabiana Sherine~Ganem Dos~Santos, Wildo~Navegantes
  Ara{\'u}jo, Maria Regina~Fernandes de~Oliveira, and Henry~Maia Peixoto.
\newblock Ten epidemiological parameters of covid-19: Use of rapid literature
  review to inform predictive models during the pandemic.
\newblock \emph{Frontiers in public health}, page 830, 2020.
\newblock \doi{10.3389/fpubh.2020.598547}.

\bibitem[MODCOV19(2020)]{modcov19review}
CNRS MODCOV19.
\newblock Covid-19 literature review, 2020.
\newblock URL \url{https://modcov19.math.cnrs.fr/veille_public/}.

\bibitem[Inserm(2020)]{insermcovidreview}
ANRS Inserm.
\newblock Covid-19 literature review, 2020.
\newblock URL \url{https://reacting.inserm.fr/covid-19/covid-19-resources-2/}.

\bibitem[Nelson and Williams(2014)]{nelson2014infectious}
Kenrad~E Nelson and Carolyn~Masters Williams.
\newblock \emph{Infectious disease epidemiology: theory and practice}.
\newblock Infectious Disease Epidemiology: Theory and Practice. Jones \&
  Bartlett Learning, 2014.
\newblock ISBN 9780763795337.

\bibitem[Chowell et~al.(2019)Chowell, Tariq, and Hyman]{chowell2019novel}
Gerardo Chowell, Amna Tariq, and James~M Hyman.
\newblock A novel sub-epidemic modeling framework for short-term forecasting
  epidemic waves.
\newblock \emph{BMC medicine}, 17\penalty0 (1):\penalty0 1--18, 2019.

\bibitem[Kermack and McKendrick(1927)]{kermack1927contribution}
William~Ogilvy Kermack and Anderson~G McKendrick.
\newblock A contribution to the mathematical theory of epidemics.
\newblock \emph{Proceedings of the royal society of london. Series A,
  Containing papers of a mathematical and physical character}, 115\penalty0
  (772):\penalty0 700--721, 1927.

\bibitem[Hazelbag et~al.(2020)Hazelbag, Dushoff, Dominic, Mthombothi, and
  Delva]{hazelbag2020calibration}
C~Marijn Hazelbag, Jonathan Dushoff, Emanuel~M Dominic, Zinhle~E Mthombothi,
  and Wim Delva.
\newblock Calibration of individual-based models to epidemiological data: A
  systematic review.
\newblock \emph{PLoS computational biology}, 16\penalty0 (5):\penalty0
  e1007893, 2020.

\bibitem[Gallagher and Baltimore(2017)]{gallagher2017comparing}
Shannon Gallagher and JSM Baltimore.
\newblock Comparing compartment and agent-based models.
\newblock 2017.

\bibitem[Nepomuceno et~al.(2019)Nepomuceno, Resende, and
  Lacerda]{nepomuceno2019survey}
EG~Nepomuceno, Denise~F Resende, and M{\'a}rcio~J Lacerda.
\newblock A survey of the individual-based model applied in biomedical and
  epidemiology.
\newblock \emph{arXiv preprint arXiv:1902.02784}, 2019.

\bibitem[Hunter et~al.(2017)Hunter, Mac~Namee, and
  Kelleher]{hunter2017taxonomy}
Elizabeth Hunter, Brian Mac~Namee, and John~D Kelleher.
\newblock A taxonomy for agent-based models in human infectious disease
  epidemiology.
\newblock \emph{Journal of Artificial Societies and Social Simulation},
  20\penalty0 (3), 2017.

\bibitem[Ferguson et~al.(2005)Ferguson, Cummings, Cauchemez, Fraser, Riley,
  Meeyai, Iamsirithaworn, and Burke]{ferguson2005strategies}
Neil~M Ferguson, Derek~AT Cummings, Simon Cauchemez, Christophe Fraser, Steven
  Riley, Aronrag Meeyai, Sopon Iamsirithaworn, and Donald~S Burke.
\newblock Strategies for containing an emerging influenza pandemic in southeast
  asia.
\newblock \emph{Nature}, 437\penalty0 (7056):\penalty0 209--214, 2005.

\bibitem[Garin et~al.(2020{\natexlab{a}})Garin, Limnios, Nicolaï, Vayatis,
  Boulant, and Dib]{github_review}
Marie Garin, Myrto Limnios, Alice Nicolaï, Nicolas Vayatis, Olivier Boulant,
  and Amir Dib.
\newblock Github covid-19 review, 2020{\natexlab{a}}.
\newblock URL \url{https://github.com/MyrtoLimnios/covid19-biblio}.

\bibitem[Garin et~al.(2020{\natexlab{b}})Garin, Limnios, Nicolaï, Vayatis,
  Boulant, and Dib]{kibana_review}
Marie Garin, Myrto Limnios, Alice Nicolaï, Nicolas Vayatis, Olivier Boulant,
  and Amir Dib.
\newblock Kibana covid-19 review, 2020{\natexlab{b}}.
\newblock URL \url{https://bit.ly/3a6tg2V}.

\bibitem[Garin et~al.(2020{\natexlab{c}})Garin, Limnios, Nicolaï, and
  Vayatis]{googlesheet_review}
Marie Garin, Myrto Limnios, Alice Nicolaï, and Nicolas Vayatis.
\newblock Google sheet covid-19 review, 2020{\natexlab{c}}.
\newblock URL \url{https://bit.ly/3dfTxOx}.

\bibitem[Roy and Karmakar(2020)]{roy2020}
Arkaprava Roy and Sayar Karmakar.
\newblock Bayesian semiparametric time varying model for count data to study
  the spread of the covid-19 cases.
\newblock \emph{arXiv preprint arXiv:2004.02281}, 2020.

\bibitem[Kraemer et~al.(2020)Kraemer, Yang, Gutierrez, Wu, Klein, Pigott,
  du~Plessis, Faria, Li, Hanage, et~al.]{kraemer2020effect}
Moritz~UG Kraemer, Chia-Hung Yang, Bernardo Gutierrez, Chieh-Hsi Wu, Brennan
  Klein, David~M Pigott, Louis du~Plessis, Nuno~R Faria, Ruoran Li, William~P
  Hanage, et~al.
\newblock The effect of human mobility and control measures on the covid-19
  epidemic in china.
\newblock \emph{Science}, 368\penalty0 (6490):\penalty0 493--497, 2020.

\bibitem[Villalobos-Arias(2020)]{villa20}
Mario Villalobos-Arias.
\newblock Using generalized logistics regression to forecast population
  infected by covid-19.
\newblock \emph{arXiv preprint arXiv:2004.02406}, 2020.

\bibitem[Verity et~al.(2020)Verity, Okell, Dorigatti, Winskill, Whittaker,
  Imai, Cuomo-Dannenburg, Thompson, Walker, Fu, et~al.]{verity2020estimates}
Robert Verity, Lucy~C Okell, Ilaria Dorigatti, Peter Winskill, Charles
  Whittaker, Natsuko Imai, Gina Cuomo-Dannenburg, Hayley Thompson, Patrick~GT
  Walker, Han Fu, et~al.
\newblock Estimates of the severity of coronavirus disease 2019: a model-based
  analysis.
\newblock \emph{The Lancet Infectious Diseases}, 20\penalty0 (6):\penalty0
  669--677, 2020.

\bibitem[Yang et~al.(2020)Yang, Zhang, Peng, Zhuge, and Hong]{yang2020rational}
Wuyue Yang, Dongyan Zhang, Liangrong Peng, Changjing Zhuge, and Liu Hong.
\newblock Rational evaluation of various epidemic models based on the covid-19
  data of china.
\newblock \emph{arXiv preprint arXiv:2003.05666}, 2020.

\bibitem[Batista(2020{\natexlab{a}})]{Batista20}
Milan Batista.
\newblock Estimation of the final size of coronavirus epidemic by the logistic
  model.
\newblock \emph{medRxiv 2020.03.11.20024901}, 2020{\natexlab{a}}.

\bibitem[COVID et~al.(2020)COVID, Murray, et~al.]{covid2020forecasting}
IHME COVID, Christopher~JL Murray, et~al.
\newblock Forecasting the impact of the first wave of the covid-19 pandemic on
  hospital demand and deaths for the usa and european economic area countries.
\newblock \emph{medRxiv:2020.04.21.20074732}, 2020.

\bibitem[Zeng et~al.(2020)Zeng, Zhang, Li, Liu, and Qiu]{zeng2020predictions}
Tianyu Zeng, Yunong Zhang, Zhenyu Li, Xiao Liu, and Binbin Qiu.
\newblock Predictions of 2019-ncov transmission ending via comprehensive
  methods.
\newblock \emph{arXiv preprint arXiv:2002.04945}, 2020.

\bibitem[Shim et~al.(2020)Shim, Tariq, Choi, Lee, and
  Chowell]{shim2020transmission}
Eunha Shim, Amna Tariq, Wongyeong Choi, Yiseul Lee, and Gerardo Chowell.
\newblock Transmission potential and severity of covid-19 in south korea.
\newblock \emph{International Journal of Infectious Diseases}, 93:\penalty0
  339--344, 2020.
\newblock ISSN 1201-9712.
\newblock \doi{https://doi.org/10.1016/j.ijid.2020.03.031}.

\bibitem[Woody et~al.(2020)Woody, Tec, Dahan, Gaither, Lachmann, Fox, Meyers,
  and Scott]{woody2020projections}
Spencer Woody, Mauricio~Garcia Tec, Maytal Dahan, Kelly Gaither, Michael
  Lachmann, Spencer Fox, Lauren~Ancel Meyers, and James~G Scott.
\newblock Projections for first-wave covid-19 deaths across the us using
  social-distancing measures derived from mobile phones.
\newblock \emph{medRxiv 2020.04.16.20068163}, 2020.

\bibitem[Roux et~al.(2020)Roux, Massonnaud, and Cr{\'e}pey]{roux2020covid}
Jonathan Roux, Cl{\'e}ment Massonnaud, and Pascal Cr{\'e}pey.
\newblock Covid-19: One-month impact of the french lockdown on the epidemic
  burden.
\newblock \emph{medRxiv 2020.04.22.20075705}, 2020.

\bibitem[Di~Domenico et~al.()Di~Domenico, Pullano, Coletti, Hens, and
  Colizza]{diexpected}
Laura Di~Domenico, Giulia Pullano, Pietro Coletti, Niel Hens, and Vittoria
  Colizza.
\newblock Expected impact of school closure and telework to mitigate covid-19
  epidemic in france.
\newblock URL
  \url{http://www.epicx-lab.com/uploads/9/6/9/4/9694133/inserm_covid-19-school-closure-french-regions_20200313.pdf}.

\bibitem[Di~Domenico et~al.(2020)Di~Domenico, Pullano, Sabbatini, Bo{\"e}lle,
  and Colizza]{di2020expected}
Laura Di~Domenico, Giulia Pullano, Chiara~E Sabbatini, Pierre-Yves Bo{\"e}lle,
  and Vittoria Colizza.
\newblock Expected impact of lockdown in ile-de-france and possible exit
  strategies.
\newblock \emph{medRxiv 2020.04.13.20063933}, 2020.

\bibitem[Jia et~al.(2020{\natexlab{a}})Jia, Ding, Liu, Liao, Li, Duan, Wang,
  and Zhang]{jia2020modeling}
Jiwei Jia, Jian Ding, Siyu Liu, Guidong Liao, Jingzhi Li, Ben Duan, Guoqing
  Wang, and Ran Zhang.
\newblock Modeling the control of covid-19: Impact of policy interventions and
  meteorological factors.
\newblock \emph{arXiv preprint arXiv:2003.02985}, 2020{\natexlab{a}}.

\bibitem[Djidjou-Demasse et~al.(2020)Djidjou-Demasse, Michalakis, Choisy,
  Sofonea, and Alizon]{djidjou2020optimal}
Ramses Djidjou-Demasse, Yannis Michalakis, Marc Choisy, Micea~T Sofonea, and
  Samuel Alizon.
\newblock Optimal covid-19 epidemic control until vaccine deployment.
\newblock \emph{medRxiv}, 2020.
\newblock \doi{10.1101/2020.04.02.20049189}.

\bibitem[Manchein et~al.(2020{\natexlab{a}})Manchein, Brugnago, da~Silva,
  Mendes, and Beims]{manchein2020strong}
Cesar Manchein, Eduardo~L Brugnago, Rafael~M da~Silva, Carlos~FO Mendes, and
  Marcus~W Beims.
\newblock Strong correlations between power-law growth of covid-19 in four
  continents and the inefficiency of soft quarantine strategies.
\newblock \emph{Chaos: An Interdisciplinary Journal of Nonlinear Science},
  30\penalty0 (4):\penalty0 041102, 2020{\natexlab{a}}.

\bibitem[Zhang and Enns(2020)]{zhangenns20}
Zhenhuan Zhang and Eva Enns.
\newblock Optimal timing and effectiveness of covid-19 outbreak responses in
  china: A modelling study.
\newblock 2020.
\newblock \doi{10.2139/ssrn.3558339}.

\bibitem[Prem et~al.(2020)Prem, Liu, Russell, Kucharski, Eggo, Davies, Flasche,
  Clifford, Pearson, Munday, et~al.]{prem2020effect}
Kiesha Prem, Yang Liu, Timothy~W Russell, Adam~J Kucharski, Rosalind~M Eggo,
  Nicholas Davies, Stefan Flasche, Samuel Clifford, Carl~AB Pearson, James~D
  Munday, et~al.
\newblock The effect of control strategies to reduce social mixing on outcomes
  of the covid-19 epidemic in wuhan, china: a modelling study.
\newblock \emph{The Lancet Public Health}, 2020.
\newblock \doi{10.1016/S2468-2667(20)30073-6}.

\bibitem[Salje et~al.(2020)Salje, Tran~Kiem, Lefrancq, Courtejoie, Bosetti,
  Paireau, Andronico, Hoze, Richet, Dubost, Le~Strat, Lessler, Bruhl, Fontanet,
  Opatowski, Bo{\"e}lle, and Cauchemez]{salje2020estimating}
Henrik Salje, C{\'e}cile Tran~Kiem, No{\'e}mie Lefrancq, No{\'e}mie Courtejoie,
  Paolo Bosetti, Juliette Paireau, Alessio Andronico, Nathana{\"e}l Hoze,
  Jehanne Richet, Claire-Lise Dubost, Yann Le~Strat, Justin Lessler,
  Daniel~Levy Bruhl, Arnaud Fontanet, Lulla Opatowski, Pierre-Yves Bo{\"e}lle,
  and Simon Cauchemez.
\newblock Estimating the burden of sars-cov-2 in france.
\newblock \emph{medRxiv}, 2020.
\newblock \doi{10.1101/2020.04.20.20072413}.

\bibitem[Thakkar et~al.(2020)Thakkar, Burstein, Klein, Schripsema, and
  Famulare]{Thakkar20}
Niket Thakkar, Roy Burstein, Daniel Klein, Jen Schripsema, and Mike Famulare.
\newblock Physical distancing is working and still needed to prevent covid-19
  resurgence in king, snohomish, and pierce counties.
\newblock 2020.
\newblock URL
  \url{https://iazpvnewgrp01.blob.core.windows.net/source/archived/Physical_distancing_working_and_still_needed_to_prevent_COVID-19_resurgence.pdf}.

\bibitem[Calafiore et~al.(2020)Calafiore, Novara, and
  Possieri]{calafiore2020modified}
Giuseppe~C. Calafiore, Carlo Novara, and Corrado Possieri.
\newblock A modified sir model for the covid-19 contagion in italy.
\newblock \emph{arXiv:2003.14391}, 2020.

\bibitem[Magal and Webb(2020)]{magal2020predicting}
Pierre Magal and Glenn Webb.
\newblock Predicting the number of reported and unreported cases for the
  covid-19 epidemic in south korea, italy, france and germany.
\newblock \emph{medRxiv}, 2020.
\newblock \doi{10.1101/2020.03.21.20040154}.

\bibitem[Gollier(2020)]{gollier2020policy}
Christian Gollier.
\newblock Policy brief : Analyse coût‐bénéfice des stratégies de
  déconfinement.
\newblock 2020.
\newblock URL
  \url{https://www.tse-fr.eu/sites/default/files/TSE/documents/doc/by/gollier/policy-brief-deconfinement-c-gollier-avril-2020.pdf}.

\bibitem[Chen et~al.(2020{\natexlab{a}})Chen, Cheng, Jiang, and
  Liu]{chen2020time}
Yu~Chen, Jin Cheng, Yu~Jiang, and Keji Liu.
\newblock A time delay dynamic system with external source for the local
  outbreak of 2019-ncov.
\newblock 2020{\natexlab{a}}.

\bibitem[Flaxman et~al.(2020)Flaxman, Mishra, Gandy, Unwin, Coupland, Mellan,
  Zhu, Berah, Eaton, Guzman, Schmit, Callizo, Team, Whittaker, Winskill, Xi,
  Ghani, Donnelly, Riley, Okell, Vollmer, Ferguson, and
  Bhatt]{flaxman2020report}
Seth Flaxman, Swapnil Mishra, Axel Gandy, H~Juliette~T Unwin, Helen Coupland,
  Thomas~A Mellan, Harrison Zhu, Tresnia Berah, Jeffrey~W Eaton, Pablo N~P
  Guzman, Nora Schmit, Lucia Callizo, Imperial College COVID-19~Response Team,
  Charles Whittaker, Peter Winskill, Xiaoyue Xi, Azra Ghani, Christl~A.
  Donnelly, Steven Riley, Lucy~C Okell, Michaela A~C Vollmer, Neil~M. Ferguson,
  and Samir Bhatt.
\newblock Report 13: Estimating the number of infections and the impact of
  non-pharmaceutical interventions on covid-19 in 11 european countries.
\newblock 2020.
\newblock URL
  \url{https://www.imperial.ac.uk/media/imperial-college/medicine/mrc-gida/2020-03-30-COVID19-Report-13.pdf}.

\bibitem[Volpert et~al.(2020)Volpert, Banerjee, and
  Petrovskii]{volpert2020quarantine}
Vitaly Volpert, Malay Banerjee, and Sergei Petrovskii.
\newblock On a quarantine model of coronavirus infection and data analysis.
\newblock \emph{Mathematical Modelling of Natural Phenomena}, 15:\penalty0 24,
  2020.

\bibitem[Kucharski et~al.(2020)Kucharski, Russell, Diamond, Liu, Edmunds, Funk,
  Eggo, Sun, Jit, Munday, et~al.]{kucharski2020early}
Adam~J Kucharski, Timothy~W Russell, Charlie Diamond, Yang Liu, John Edmunds,
  Sebastian Funk, Rosalind~M Eggo, Fiona Sun, Mark Jit, James~D Munday, et~al.
\newblock Early dynamics of transmission and control of covid-19: a
  mathematical modelling study.
\newblock \emph{The lancet infectious diseases}, 2020.
\newblock \doi{D10.1016/S1473-3099(20)30144-4}.

\bibitem[Chen and Qiu(2020)]{chen2020scenario}
Xiaohui Chen and Ziyi Qiu.
\newblock Scenario analysis of non-pharmaceutical interventions on global
  covid-19 transmissions.
\newblock \emph{arXiv:2004.04529}, 2020.

\bibitem[Zheng(2020)]{zheng2020total}
Wenjie Zheng.
\newblock Total variation regularization for compartmental epidemic models with
  time-varying dynamics.
\newblock \emph{arXiv:2004.00412}, 2020.

\bibitem[Zhigljavsky et~al.(2020)Zhigljavsky, Whitaker, Fesenko, Kremnizer,
  Noonan, Harper, Gillard, Woolley, Gartner, Grimsley,
  et~al.]{zhigljavsky2020generic}
Anatoly Zhigljavsky, Roger Whitaker, Ivan Fesenko, Kobi Kremnizer, Jack Noonan,
  Paul Harper, Jonathan Gillard, Thomas Woolley, Daniel Gartner, Jasmine
  Grimsley, et~al.
\newblock Generic probabilistic modelling and non-homogeneity issues for the uk
  epidemic of covid-19.
\newblock \emph{arXiv:2004.01991}, 2020.

\bibitem[Weissman et~al.(2020)Weissman, Crane-Droesch, Chivers, Luong, Hanish,
  Levy, Lubken, Becker, Draugelis, Anesi, et~al.]{weissman2020locally}
Gary~E Weissman, Andrew Crane-Droesch, Corey Chivers, ThaiBinh Luong, Asaf
  Hanish, Michael~Z Levy, Jason Lubken, Michael Becker, Michael~E Draugelis,
  George~L Anesi, et~al.
\newblock Locally informed simulation to predict hospital capacity needs during
  the covid-19 pandemic.
\newblock \emph{Annals of internal medicine}, 2020.

\bibitem[Khadilkar et~al.(2020)Khadilkar, Ganu, and
  Seetharam]{khadilkar2020optimising}
Harshad Khadilkar, Tanuja Ganu, and Deva~P Seetharam.
\newblock Optimising lockdown policies for epidemic control using reinforcement
  learning.
\newblock \emph{arXiv:2003.14093}, 2020.

\bibitem[Simha et~al.(2020)Simha, Prasad, and Narayana]{simha2020simple}
Ashutosh Simha, R~Venkatesha Prasad, and Sujay Narayana.
\newblock A simple stochastic sir model for covid 19 infection dynamics for
  karnataka: Learning from europe.
\newblock \emph{arXiv:2003.11920}, 2020.

\bibitem[Wu et~al.(2020{\natexlab{a}})Wu, Leung, Bushman, Kishore, Niehus,
  de~Salazar, Cowling, Lipsitch, and Leung]{wu2020estimating}
Joseph~T Wu, Kathy Leung, Mary Bushman, Nishant Kishore, Rene Niehus, Pablo~M
  de~Salazar, Benjamin~J Cowling, Marc Lipsitch, and Gabriel~M Leung.
\newblock Estimating clinical severity of covid-19 from the transmission
  dynamics in wuhan, china.
\newblock \emph{Nature Medicine}, 26\penalty0 (4):\penalty0 506--510,
  2020{\natexlab{a}}.

\bibitem[Roques et~al.(2020)Roques, Klein, Papaix, and
  Soubeyrand]{roques2020modele}
Lionel Roques, Etienne Klein, Julien Papaix, and Samuel Soubeyrand.
\newblock \emph{Modèle SIR mécanistico-statistique pour l'estimation du
  nombre d'infectés et du taux de mortalité par COVID-19}.
\newblock PhD thesis, INRAE, 2020.

\bibitem[Wu et~al.(2020{\natexlab{b}})Wu, Leung, and Leung]{wu2020nowcasting}
Joseph~T Wu, Kathy Leung, and Gabriel~M Leung.
\newblock Nowcasting and forecasting the potential domestic and international
  spread of the 2019-ncov outbreak originating in wuhan, china: a modelling
  study.
\newblock \emph{The Lancet}, 395\penalty0 (10225):\penalty0 689--697,
  2020{\natexlab{b}}.

\bibitem[Chikina and Pegden(2020)]{chikina2020modeling}
Maria Chikina and Wesley Pegden.
\newblock Modeling strict age-targeted mitigation strategies for covid-19.
\newblock \emph{arXiv:2004.04144}, 2020.

\bibitem[Massonnaud et~al.(2020)Massonnaud, Roux, and
  Cr{\'e}pey]{massonnaud2020covid}
Cl{\'e}ment Massonnaud, Jonathan Roux, and Pascal Cr{\'e}pey.
\newblock Covid-19: Forecasting short term hospital needs in france.
\newblock \emph{medRxiv:2020.03.16.20036939}, 2020.

\bibitem[Zhang et~al.(2020{\natexlab{a}})Zhang, Dong, Zhang, Chen, Yao, and
  Han]{zhang2020predicting}
Jiang Zhang, Lei Dong, Yanbo Zhang, Xinyue Chen, Guiqing Yao, and Zhangang Han.
\newblock Predicting the spread of the covid-19 across cities in china with
  population migration and policy intervention.
\newblock 2020{\natexlab{a}}.
\newblock \doi{10.2139/ssrn.3559554}.

\bibitem[Li et~al.(2020{\natexlab{a}})Li, Pei, Chen, Song, Zhang, Yang, and
  Shaman]{li2020substantial}
Ruiyun Li, Sen Pei, Bin Chen, Yimeng Song, Tao Zhang, Wan Yang, and Jeffrey
  Shaman.
\newblock Substantial undocumented infection facilitates the rapid
  dissemination of novel coronavirus (sars-cov-2).
\newblock \emph{Science}, 368\penalty0 (6490):\penalty0 489--493,
  2020{\natexlab{a}}.

\bibitem[Magdon-Ismail(2020)]{magdon2020machine}
Malik Magdon-Ismail.
\newblock Machine learning the phenomenology of covid-19 from early infection
  dynamics.
\newblock \emph{arXiv:2003.07602}, 2020.

\bibitem[Wood et~al.(2020)Wood, Warrington, Naderiparizi, Weilbach, Masrani,
  Harvey, Scibior, Beronov, and Nasseri]{wood2020planning}
Frank Wood, Andrew Warrington, Saeid Naderiparizi, Christian Weilbach, Vaden
  Masrani, William Harvey, Adam Scibior, Boyan Beronov, and Ali Nasseri.
\newblock Planning as inference in epidemiological models.
\newblock \emph{arXiv:2003.13221}, 2020.

\bibitem[Chen et~al.(2020{\natexlab{b}})Chen, Lu, Chang, and Liu]{liutime}
Yi-Cheng Chen, Ping-En Lu, Cheng-Shang Chang, and Tzu-Hsuan Liu.
\newblock A time-dependent sir model for covid-19 with undetectable infected
  persons.
\newblock \emph{IEEE Transactions on Network Science and Engineering},
  7\penalty0 (4):\penalty0 3279–3294, 2020{\natexlab{b}}.
\newblock ISSN 2334-329X.
\newblock \doi{10.1109/tnse.2020.3024723}.

\bibitem[Liu et~al.(2020{\natexlab{a}})Liu, Magal, Seydi, and
  Webb]{liu2020predicting}
Zhihua Liu, Pierre Magal, Ousmane Seydi, and Glenn Webb.
\newblock Predicting the cumulative number of cases for the covid-19 epidemic
  in china from early data.
\newblock \emph{arXiv:2002.12298}, 2020{\natexlab{a}}.

\bibitem[Liu et~al.(2020{\natexlab{b}})Liu, Magal, Seydi, and
  Webb]{liu2020understanding}
Zhihua Liu, Pierre Magal, Ousmane Seydi, and Glenn Webb.
\newblock Understanding unreported cases in the covid-19 epidemic outbreak in
  wuhan, china, and the importance of major public health interventions.
\newblock \emph{Biology}, 9\penalty0 (3):\penalty0 50, 2020{\natexlab{b}}.

\bibitem[Yanev et~al.(2020)Yanev, Stoimenova, and
  Atanasov]{yanev2020stochastic}
Nikolay~M Yanev, Vessela~K Stoimenova, and Dimitar~V Atanasov.
\newblock Stochastic modeling and estimation of covid-19 population dynamics.
\newblock \emph{arXiv preprint arXiv:2004.00941}, 2020.

\bibitem[Hurd(2020)]{hurd2020covid}
T.~R. Hurd.
\newblock Covid-19: Analytics of contagion on inhomogeneous random social
  networks.
\newblock \emph{arXiv preprint arXiv:2004.02779}, 2020.

\bibitem[Karin et~al.(2020)Karin, Bar-On, Milo, Katzir, Mayo, Korem, Dudovich,
  Yashiv, Zehavi, Davidovich, et~al.]{karin2020adaptive}
Omer Karin, Yinon~M Bar-On, Tomer Milo, Itay Katzir, Avi Mayo, Yael Korem, Boaz
  Dudovich, Eran Yashiv, Amos~J Zehavi, Nadav Davidovich, et~al.
\newblock Adaptive cyclic exit strategies from lockdown to suppress covid-19
  and allow economic activity.
\newblock \emph{medRxiv 2020.04.04.20053579}, 2020.

\bibitem[Ferguson et~al.(2020)Ferguson, Laydon, Nedjati~Gilani, Imai, Ainslie,
  Baguelin, Bhatia, Boonyasiri, Cucunuba~Perez, Cuomo-Dannenburg,
  et~al.]{ferguson2020report}
Neil Ferguson, Daniel Laydon, Gemma Nedjati~Gilani, Natsuko Imai, Kylie
  Ainslie, Marc Baguelin, Sangeeta Bhatia, Adhiratha Boonyasiri, ZULMA
  Cucunuba~Perez, Gina Cuomo-Dannenburg, et~al.
\newblock Impact of non-pharmaceutical interventions (npis) to reduce covid-19
  mortality and healthcare demand.
\newblock Technical report, Imperial College London, 2020.
\newblock URL \url{http://hdl.handle.net/10044/1/77482}.

\bibitem[team(2020)]{gleam2020modeling}
GLEAM team.
\newblock Modeling of covid-19 epidemic in the united states.
\newblock Technical report, Northeastern University, Boston, 2020.
\newblock URL
  \url{https://uploads-ssl.webflow.com/58e6558acc00ee8e4536c1f5/5e8bab44f5baae4c1c2a75d2_GLEAM_web.pdf}.

\bibitem[Anzai et~al.(2020)Anzai, Kobayashi, Linton, Kinoshita, Hayashi,
  Suzuki, Yang, Jung, Miyama, Akhmetzhanov, et~al.]{anzai2020assessing}
Asami Anzai, Tetsuro Kobayashi, Natalie~M Linton, Ryo Kinoshita, Katsuma
  Hayashi, Ayako Suzuki, Yichi Yang, Sung-mok Jung, Takeshi Miyama, Andrei~R
  Akhmetzhanov, et~al.
\newblock Assessing the impact of reduced travel on exportation dynamics of
  novel coronavirus infection (covid-19).
\newblock \emph{Journal of clinical medicine}, 9\penalty0 (2):\penalty0 601,
  2020.

\bibitem[Das(2020{\natexlab{a}})]{das2020}
Sourish Das.
\newblock Prediction of covid-19 disease progression in india : Under the
  effect of national lockdown.
\newblock \emph{arXiv preprint arXiv:2004.03147}, 2020{\natexlab{a}}.

\bibitem[Evgeniou et~al.(2020)Evgeniou, Fekom, Ovchinnikov, Porcher, Pouchol,
  and Vayatis]{evgeniou2020epidemic}
Theodoros Evgeniou, Mathilde Fekom, Anton Ovchinnikov, Raphael Porcher, Camille
  Pouchol, and Nicolas Vayatis.
\newblock Epidemic models for personalised covid-19 isolation and exit policies
  using clinical risk predictions.
\newblock \emph{Available at SSRN 3588401}, 2020.

\bibitem[Piguillem and Shi(2020)]{piguillem2020optimal}
Facundo Piguillem and Liyan Shi.
\newblock Optimal covid-19 quarantine and testing policies.
\newblock Technical report, Einaudi Institute for Economics and Finance (EIEF),
  2020.
\newblock URL \url{http://www.eief.it/eief/images/WP_20.04.pdf}.

\bibitem[Victor(2020)]{victor2020mathematical}
Alexander Victor.
\newblock Mathematical predictions for covid-19 as a global pandemic.
\newblock \emph{Available at SSRN 3555879}, 2020.

\bibitem[Jia et~al.(2020{\natexlab{b}})Jia, Ding, Liu, Liao, Li, Duan, Wang,
  and Zhang]{Jiwei20}
Jiwei Jia, Jian Ding, Siyu Liu, Guidong Liao, Jingzhi Li, Ben Duan, Guoqing
  Wang, and Ran Zhang.
\newblock Modeling the control of covid-19: impact of policy interventions and
  meteorological factors.
\newblock \emph{Electronic Journal of Differential Equations}, 2020:\penalty0
  1--24, 03 2020{\natexlab{b}}.

\bibitem[Chinazzi et~al.(2020)Chinazzi, Davis, Ajelli, Gioannini, Litvinova,
  Merler, y~Piontti, Mu, Rossi, Sun, et~al.]{chinazzi2020effect}
Matteo Chinazzi, Jessica~T Davis, Marco Ajelli, Corrado Gioannini, Maria
  Litvinova, Stefano Merler, Ana~Pastore y~Piontti, Kunpeng Mu, Luca Rossi,
  Kaiyuan Sun, et~al.
\newblock The effect of travel restrictions on the spread of the 2019 novel
  coronavirus (covid-19) outbreak.
\newblock \emph{Science}, 368\penalty0 (6489):\penalty0 395--400, 2020.

\bibitem[Alvarez et~al.(2020)Alvarez, Argente, and Lippi]{alvarez2020simple}
Fernando~E Alvarez, David Argente, and Francesco Lippi.
\newblock A simple planning problem for covid-19 lockdown.
\newblock Technical report, National Bureau of Economic Research, 2020.
\newblock URL
  \url{https://bfi.uchicago.edu/wp-content/uploads/BFI_WP_202034.pdf}.

\bibitem[Toda(2020)]{toda20}
Alexis~Akira Toda.
\newblock Susceptible-infected-recovered (sir) dynamics of covid-19 and
  economic impact.
\newblock \emph{arXiv preprint arXiv:2003.11221}, 2020.

\bibitem[Manchein et~al.(2020{\natexlab{b}})Manchein, Brugnago, da~Silva,
  Mendes, and Beims]{manchein20}
Cesar Manchein, Eduardo~L. Brugnago, Rafael~M. da~Silva, Carlos F.~O. Mendes,
  and Marcus~W. Beims.
\newblock Strong correlations between power-law growth of covid-19 in four
  continents and the inefficiency of soft quarantine strategies.
\newblock \emph{arXiv preprint arXiv:2004.00044}, 2020{\natexlab{b}}.

\bibitem[Li et~al.(2020{\natexlab{b}})Li, Wang, Gilmour, Wang, Yoneoka, Wang,
  You, Gu, Hao, Peng, Du, Xu, and Hao]{li2020estimation}
Jinghua Li, Yijing Wang, Stuart Gilmour, Mengying Wang, Daisuke Yoneoka, Ying
  Wang, Xinyi You, Jing Gu, Chun Hao, Liping Peng, Zhicheng Du, Dong~(Roman)
  Xu, and Yuantao Hao.
\newblock Estimation of the epidemic properties of the 2019 novel coronavirus:
  A mathematical modeling study.
\newblock \emph{Available at SSRN 3542150}, 2020{\natexlab{b}}.

\bibitem[Zhang et~al.(2020{\natexlab{b}})Zhang, Litvinova, Wang, Wang, Deng,
  Chen, Li, Zheng, Yi, Chen, et~al.]{zhang2020evolving}
Juanjuan Zhang, Maria Litvinova, Wei Wang, Yan Wang, Xiaowei Deng, Xinghui
  Chen, Mei Li, Wen Zheng, Lan Yi, Xinhua Chen, et~al.
\newblock Evolving epidemiology and transmission dynamics of coronavirus
  disease 2019 outside hubei province, china: a descriptive and modelling
  study.
\newblock \emph{The Lancet Infectious Diseases}, 2020{\natexlab{b}}.

\bibitem[Piccolomiini and Zama(2020)]{piccolomiini2020monitoring}
Elena~Loli Piccolomiini and Fabiana Zama.
\newblock Monitoring italian covid-19 spread by an adaptive seird model.
\newblock \emph{medRxiv 2020.04.03.20049734}, 2020.

\bibitem[Dandekar and Barbastathis(2020)]{dandekar2020quantifying}
Raj Dandekar and George Barbastathis.
\newblock Quantifying the effect of quarantine control in covid-19 infectious
  spread using machine learning.
\newblock \emph{medRxiv 2020.04.03.20052084}, 2020.

\bibitem[De~Falco et~al.(2020)De~Falco, Della~Cioppa, Scafuri, and
  Tarantino]{de2020coronavirus}
I~De~Falco, A~Della~Cioppa, U~Scafuri, and E~Tarantino.
\newblock Coronavirus covid-19 spreading in italy: optimizing an
  epidemiological model with dynamic social distancing through differential
  evolution.
\newblock \emph{arXiv preprint arXiv:2004.00553}, 2020.

\bibitem[Nadim et~al.(2020)Nadim, Ghosh, and Chattopadhyay]{nadim2020short}
Sk~Shahid Nadim, Indrajit Ghosh, and Joydev Chattopadhyay.
\newblock Short-term predictions and prevention strategies for covid-2019: A
  model based study.
\newblock \emph{arXiv preprint arXiv:2003.08150}, 2020.

\bibitem[Das(2020{\natexlab{b}})]{das2020prediction}
Sourish Das.
\newblock Prediction of covid-19 disease progression in india: Under the effect
  of national lockdown.
\newblock \emph{arXiv preprint arXiv:2004.03147}, 2020{\natexlab{b}}.

\bibitem[Zhang et~al.(2020{\natexlab{c}})Zhang, Litvinova, Wang, Wang, Deng,
  Chen, Li, Zheng, Yi, Chen, Wu, Liang, Wang, Yang, Sun, Longini, Halloran, Wu,
  Cowling, Merler, Viboud, Vespignani, Ajelli, and Yu]{ZHANGlit2020}
Juanjuan Zhang, Maria Litvinova, Wei Wang, Yan Wang, Xiaowei Deng, Xinghui
  Chen, Mei Li, Wen Zheng, Lan Yi, Xinhua Chen, Qianhui Wu, Yuxia Liang, Xiling
  Wang, Juan Yang, Kaiyuan Sun, Ira~M Longini, M~Elizabeth Halloran, Peng Wu,
  Benjamin~J Cowling, Stefano Merler, Cecile Viboud, Alessandro Vespignani,
  Marco Ajelli, and Hongjie Yu.
\newblock Evolving epidemiology and transmission dynamics of coronavirus
  disease 2019 outside hubei province, china: a descriptive and modelling
  study.
\newblock \emph{The Lancet Infectious Diseases}, 2020{\natexlab{c}}.
\newblock ISSN 1473-3099.
\newblock \doi{https://doi.org/10.1016/S1473-3099(20)30230-9}.

\bibitem[Linton et~al.(2020)Linton, Kobayashi, Yang, Hayashi, Akhmetzhanov,
  Jung, Yuan, Kinoshita, and Nishiura]{linton2020incubation}
Natalie~M Linton, Tetsuro Kobayashi, Yichi Yang, Katsuma Hayashi, Andrei~R
  Akhmetzhanov, Sung-mok Jung, Baoyin Yuan, Ryo Kinoshita, and Hiroshi
  Nishiura.
\newblock Incubation period and other epidemiological characteristics of 2019
  novel coronavirus infections with right truncation: a statistical analysis of
  publicly available case data.
\newblock \emph{Journal of clinical medicine}, 9\penalty0 (2):\penalty0 538,
  2020.

\bibitem[Leung et~al.(2020)Leung, Wu, Liu, and Leung]{leung2020first}
Kathy Leung, Joseph~T Wu, Di~Liu, and Gabriel~M Leung.
\newblock First-wave covid-19 transmissibility and severity in china outside
  hubei after control measures, and second-wave scenario planning: a modelling
  impact assessment.
\newblock \emph{The Lancet}, 2020.

\bibitem[He et~al.(2020)He, Lau, Wu, Deng, Wang, Hao, Lau, Wong, Guan, Tan,
  et~al.]{he2020temporal}
Xi~He, Eric~HY Lau, Peng Wu, Xilong Deng, Jian Wang, Xinxin Hao, Yiu~Chung Lau,
  Jessica~Y Wong, Yujuan Guan, Xinghua Tan, et~al.
\newblock Temporal dynamics in viral shedding and transmissibility of covid-19.
\newblock \emph{Nature medicine}, pages 1--4, 2020.

\bibitem[Gilbert et~al.(2020)Gilbert, Pullano, Pinotti, Valdano, Poletto,
  Bo{\"e}lle, d'Ortenzio, Yazdanpanah, Eholie, Altmann,
  et~al.]{gilbert2020preparedness}
Marius Gilbert, Giulia Pullano, Francesco Pinotti, Eugenio Valdano, Chiara
  Poletto, Pierre-Yves Bo{\"e}lle, Eric d'Ortenzio, Yazdan Yazdanpanah,
  Serge~Paul Eholie, Mathias Altmann, et~al.
\newblock Preparedness and vulnerability of african countries against
  importations of covid-19: a modelling study.
\newblock \emph{The Lancet}, 395\penalty0 (10227):\penalty0 871--877, 2020.

\bibitem[Bayham and Fenichel(2020)]{bayham2020impact}
Jude Bayham and Eli~P Fenichel.
\newblock The impact of school closure for covid-19 on the us healthcare
  workforce and the net mortality effects.
\newblock \emph{Available at SSRN 3555259}, 2020.

\bibitem[Luo(2020)]{luo2020monitor}
Jianxi Luo.
\newblock Predictive monitoring of covid-19.
\newblock Technical report, SUTD Data-Driven Innovation Lab, 2020.
\newblock URL
  \url{https://www.newsbeast.gr/files/1/2020/05/COVID19PredictionPaper.pdf}.

\bibitem[Batista(2020{\natexlab{b}})]{Batista20_2}
Milan Batista.
\newblock Estimation of the final size of coronavirus epidemic by the sir
  model.
\newblock \emph{ResearchGate}, 2020{\natexlab{b}}.
\newblock URL
  \url{https://www.researchgate.net/publication/339311383_Estimation_of_the_final_size_of_the_coronavirus_epidemic_by_the_SIR_model}.

\bibitem[Fong et~al.(2020)Fong, Li, Dey, Crespo, and Herrera-Viedma]{fong2020}
Simon~James Fong, Gloria Li, Nilanjan Dey, Ruben~Gonzalez Crespo, and Enrique
  Herrera-Viedma.
\newblock Composite monte carlo decision making under high uncertainty of novel
  coronavirus epidemic using hybridized deep learning and fuzzy rule induction.
\newblock \emph{arXiv preprint arXiv:2003.09868}, 2020.

\bibitem[Adiga et~al.(2020{\natexlab{a}})Adiga, Venkatramanan, Schlitt,
  Peddireddy, Dickerman, Bura, Warren, Klahn, Mao, Xie,
  et~al.]{adiga2020spread}
Aniruddha Adiga, Srinivasan Venkatramanan, James Schlitt, Akhil Peddireddy,
  Allan Dickerman, Andrei Bura, Andrew Warren, Brian~D Klahn, Chunhong Mao,
  Dawen Xie, et~al.
\newblock Evaluating the impact of international airline suspensions on the
  early global spread of covid-19.
\newblock \emph{medRxiv 2020.02.20.20025882}, 2020{\natexlab{a}}.

\bibitem[Adiga et~al.(2020{\natexlab{b}})Adiga, Venkatramanan, Peddireddy,
  Telionis, Dickerman, Wilson, Bura, Warren, Vullikanti, Klahn,
  et~al.]{adiga2020evaluating}
Aniruddha Adiga, Srinivasan Venkatramanan, Akhil Peddireddy, Alex Telionis,
  Allan Dickerman, Amanda Wilson, Andrei Bura, Andrew Warren, Anil Vullikanti,
  Brian~D Klahn, et~al.
\newblock Evaluating the impact of international airline suspensions on
  covid-19 direct importation risk.
\newblock \emph{medRxiv available at
  https://www.medrxiv.org/content/medrxiv/early/2020/02/23/2020.02.20.20025882.full.pdf},
  2020{\natexlab{b}}.

\end{thebibliography}

\end{document}